\documentclass[aps,10pt,reprint,groupedaddress,superscriptaddressm, prx, nofootinbib]{revtex4-2}
\usepackage[usenames,dvipsnames]{xcolor}
\usepackage{amssymb}
\usepackage{graphicx}
\usepackage{amsmath}
\usepackage{bbm}
\usepackage[bookmarks=true,colorlinks,citecolor=blue,urlcolor=blue]{hyperref}
\usepackage{braket}
\usepackage{babel}
\usepackage{blindtext}
\usepackage[compat=1.1.0]{tikz-feynman}
\usepackage[normalem]{ulem}
\usepackage{physics}
\usepackage{mathrsfs}
\usepackage{footmisc}
\usepackage{booktabs}

\usepackage{mathtools}

\usepackage{dsfont}

\definecolor{mypurp}{rgb}{0.35, 0, 0.7}

\usepackage{amsthm}
\usepackage{txfonts}

\usepackage{soul}

\theoremstyle{definition}

\newcommand{\tTC}
{\text{TC}}

\usepackage{lineno}
\usepackage{etoolbox}

\begin{document}

\newcommand{\thetitle}{Critical behavior of Fredenhagen-Marcu string order parameters\\ at topological phase transitions with emergent higher-form symmetries}

\title{\thetitle}

\newcommand{\TUM}{\affiliation{Technical University of Munich, TUM School of Natural Sciences, Physics Department, 85748 Garching, Germany}}
\newcommand{\MCQST}{\affiliation{Munich Center for Quantum Science and Technology (MCQST), Schellingstr. 4, 80799 M{\"u}nchen, Germany}}
\author{Wen-Tao Xu, Frank Pollmann and Michael Knap}
\affiliation{Technical University of Munich, TUM School of Natural Sciences, Physics Department, 85748 Garching, Germany}
\affiliation{Munich Center for Quantum Science and Technology (MCQST), Schellingstr. 4, 80799 M{\"u}nchen, Germany}

\begin{abstract}

A nonlocal string order parameter detecting topological order and deconfinement has been proposed by Fredenhagen and Marcu (FM). However, due to the lack of exact internal symmetries for lattice models and the nonlinear dependence of the FM string order parameter on ground states, it is \textit{a priori} not guaranteed that it is a genuine order parameter for topological phase transitions. In this work, we find that the FM string order parameter exhibits universal scaling behavior near critical points of charge condensation transitions, by directly evaluating the FM string order parameter in the infinite string-length limit using infinite Projected Entangled Pair States (iPEPS) for the toric code in a magnetic field. Our results thus demonstrate that the FM string order parameter represents a quantitatively well-behaved order parameter. 
We find that only in the presence of an \emph{emergent} 1-form symmetry the corresponding FM string order parameter can faithfully detect topological transitions.  

\end{abstract}
\maketitle

\begin{figure}
    \centering
    \includegraphics{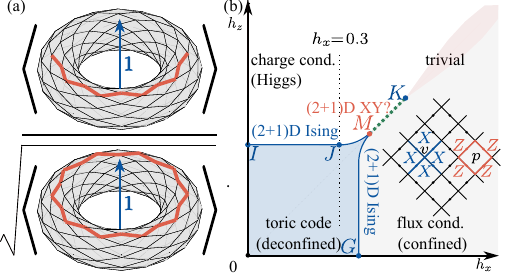}
    \caption{\textbf{FM string order parameter and phase diagram.} (a) Schematic of the FM string order parameter associated with charge excitations, where black lines indicate the underlying lattice on a torus, red lines are a string and a loop of $Z$ operators, respectively, and $\pmb{1}$ is the trivial anyon flux penetrating through the torus. (b) Phase diagram of the (2+1)D toric code model in a field, including a toric code phase (blue region), a duality symmetry breaking phase (green dotted line along $MK$) and a trivial phase (white+red+gray regions). The trivial phase is separated into three regions by emergent 1-form symmetries: the charge condensation region (white) with the emergent Wilson loop symmetry and the flux condensation region (gray) with the emergent 't Hooft loop symmetry, as well as a region without any emergent 1-form symmetry (red). Boundaries of regions in the trivial phase are shown schematically. 
    Inset: definition of the vertex and the plaquette operators on a square lattice. 
     }
    \label{Figure_1}
\end{figure}

%
%
Since the discovery of the fractional quantum Hall effect~\cite{FQHE_1982,Laughlin_1983}, topological phases of matter have been intensively explored. Exactly solvable models have been constructed~\cite{kitaev_2002,Kitaev_2006,String_net_2005}, and a mathematical framework for classifying topological phases of matter has been developed~\cite{Theory_bosonic_topo_2015,Zoo_2017}. Recently, various topologically ordered states have been experimentally investigated with quantum computers and simulators as well~\cite{TC_quantum_computer_2021,Ruben_science_2021, Iqbal_2023}. Unlike conventional phases of matter, topological phases cannot be characterized by simple local order parameters. Instead non-local string order parameters are required, for example, as proposed by Fredenhagen and Marcu (FM) in the context of the $\mathbb{Z}_2$ gauge-Higgs model~\cite{FM_1983,FM_1986,FM_PRL_1986,FM_NPB_1988}. Recently, this FM string order parameter has also been used to detect $\mathbb{Z}_2$ topological order in a quantum dimer model realized by the Rydberg quantum simulator~\cite{Ruben_PRX,Ruben_science_2021}. The construction of the FM order parameter is based on the following idea. In the presence of an exact Wilson loop symmetry, which is an exact 1-form symmetry~\cite{Zohar_2004,Hastings_and_Wen_2005,NUSSINOV_2009,High_form_Kapistin_2015,High_form_wen_2019,McGreevy_2023}, the charge condensation transition can be detected using a string order parameter carrying charges at the two ends. However, in the absence of the exact 1-form Wilson loop symmetry, the expectation value of the string order parameter decays exponentially to zero with the string length. Therefore, it cannot be used to detect phase transitions anymore. This problem is circumvented by the FM string order parameter in which the expectation value of the string operator is divided by the square root of the expectation value of a Wilson loop operator of doubled length, such that the exponential decay caused by explicitly violating the exact 1-form Wilson loop symmetry is canceled, see Fig.~\ref{Figure_1}a.

In contrast to the usual definition of the order parameters for spontaneous symmetry-breaking phases and symmetry-protected topological phases, the FM order parameter is not defined based on the exact symmetry of the model, and the FM string order parameter is a non-linear function of the ground states. Therefore, it is not  guaranteed that the FM string order parameter is quantitatively well-behaved, i.e., one cannot ensure that the FM string order parameter is smooth in a gapped phase and continuous when crossing a second-order phase transition. Even more so, it is unclear whether the FM string order parameter exhibits critical behavior, as usual order parameters do, in the vicinity of a quantum critical point. 

In this work, we study properties of the FM string order parameter of the toric code model in a field and elucidate the role of emergent higher-form symmetries. We evaluate the FM string order parameter associated with charge excitations in the limit of an infinitely long string using transfer matrices of infinite Projected Entangled Pair States (iPEPS). 
Technically, we approximate ground states of the model using variational iPEPS optimization~\cite{Laurens_gradient,PEPS_optimization_corboz}, in which energy gradients are calculated by automatic differentiation~\cite{Liao_AD_2019}.
We find that near the topological transition between the toric code and the Higgs region, Fig.~\ref{Figure_1}b, the FM string order parameter exhibits a critical exponent of the $(2+1)$D Ising* universality class~\cite{torus_CFT_spec_2016}. In the presence of the electric-magnetic (EM) duality symmetry, the phase transition has been previously studied with a local order parameter~\cite{Self_dual_critically_Ising,Multicritical_2022,Machine_learning_TC_SD_2023}. We show that the FM string order parameter exhibits critical scaling in the vicinity of this phase transition but with a new critical exponent, which is different from that of the local order parameter. The universality of this multi-critical point has been recently under debate~\cite{Self_dual_critically_Ising, Multicritical_2022, Machine_learning_TC_SD_2023,bonati_2024_comment} because although the numerical results from the scaling of local operators are consistent with the multi-critical point belonging to the XY* universality class, the semionic statistics between charge and flux is not included in the XY* field theory in an obvious way.  We find the critical exponent from the FM string order parameter to be compatible with the $(2+1)$D XY* universality class as well. However, due to the comparatively large numerical uncertainty this does not rule out other field theoretic descriptions of the multi-critical point. In general, our results show that despite being a nonlinear function of the ground state, the FM string order parameter exhibits universal scaling behavior near critical points, suggesting that it is a quantitatively well-behaved order parameter for topological phase transitions accompanied by charge condensation. 

We emphasize that the FM string order parameter associated with charge excitations can only be used to reliably detect topological phase transitions when the ground states possess an \emph{emergent} 1-form Wilson loop symmetry in the infrared (blue and white shaded regions in Fig.~\ref{Figure_1}b). 
Thus, some prior knowledge of the underlying higher-form symmetries are needed when applying the FM string order parameter to a model with an unknown phase diagram. 
{For instance,} in the flux condensation region (gray region in ~\ref{Figure_1}b) in which there is no emergent 1-form Wilson loop symmetry, the FM order parameter can be a discontinuous function (i.e., it jumps from zero to a finite value) of ground states even in the absence of phase transitions. Moreover, in the absence of an emergent 1-form symmetry we find the FM order parameter to be numerically unstable for variational iPEPS, see supplemental materials~\cite{appendix} (similar observations were found in Monte Carlo simulations~\cite{FM_1986,linsel2024percolation}). 
In particular, a dual FM string order parameter associated with magnetic flux is required to detect the topological flux condensation transition where an emergent 1-form 't Hooft loop symmetry exists [blue and gray shaded region in Fig.~\ref{Figure_1}b].

\ \\ \ \\
{\large\textbf{Results}}\\
\textbf{Toric code model and the FM string order parameter.}
We consider the toric code model on a square lattice in a field:
\begin{equation}\label{TC_Hamiltonian}
H_{\text{TC}}(h_x,h_z)=-\sum_{v}A_v-\sum_{p}B_p-h_x\sum_{e}X_e-h_z\sum_{e} Z_e,
\end{equation}
where $A_v=\prod_{e\in v}X_e$ and $B_p=\prod_{e\in p} Z_e$ are the vertex and plaquette operators, as shown in the inset of Fig.~\ref{Figure_1}b, and $X_e$ and $Z_e$ are Pauli matrices defined on the edges $e$ of the lattice. The phase diagram of the model in Fig.~\ref{Figure_1}b consists of a toric code phase at weak fields and a trivial phase at strong fields~\cite{Youjin_TC_phase_diagram_2011,TC_phase_diagram_expansion_2009,TC_XYZ_2011}. 

Recent work has introduced higher-from symmetries to describe topological order~\cite{Zohar_2004,Hastings_and_Wen_2005,NUSSINOV_2009,High_form_Kapistin_2015,High_form_wen_2019,McGreevy_2023}. 
For certain choices of the magnetic fields, the toric code model in Eq.~\eqref{TC_Hamiltonian} has exact 1-form symmetries that commute with the Hamiltonian: (i) for $h_x=0$, the 1-form Wilson loop symmetry commutes with $H_{\tTC}(0,h_z)$, $[\prod_{e\in L}Z_e,H_{\tTC}(0,h_z)]=0$, where $L$ is a closed loop on the primal lattice and (ii) for $h_z=0$ the 1-form 't Hooft loop symmetry commutes with $H_{\tTC}(h_x,0)$, $[\prod_{e\in \hat{L}}X_e,H_{\tTC}(h_x,0)]=0$, where $\hat L$ is a closed loop on the dual lattice.\footnote{On a torus, the 1-form symmetries include both contractible and  non-contractible loop operators, while on a sphere the 1-form symmetries only consist of contractible loop operators.} Crucially, even when the exact symmetries are explicitly broken, higher-form symmetries can still be emergent at low energies, capturing the robustness of topological order. Concretely, both of the 1-form symmetries are emergent in the toric code phase, see Fig.~\ref{Figure_1}b. In this formalism, topological order can be interpreted as a spontaneous breaking of such emergent symmetries, drawing analogies to conventional symmetry breaking phases~\cite{High_form_Kapistin_2015,High_form_wen_2019,McGreevy_2023,Wen_emergent_high_form_2023,my_footnote}.
%
Furthermore, the trivial phase can be separated into several regions by emergent 1-form symmetries~\cite{Self_dual_critically_Ising,Wen_emergent_high_form_2023}.
The charge condensation region (white area in Fig.~\ref{Figure_1}b), also known as the Higgs region, only has the emergent 1-form Wilson loop symmetry. The flux condensation region (gray area in Fig.~\ref{Figure_1}b), also known as the confined region, only has the emergent 1-form 't Hooft loop symmetry. The red area in Fig.~\ref{Figure_1}b does not possess any of the two emergent 1-form symmetries. 

As the toric code model in Eq.~\eqref{TC_Hamiltonian} has the exact 1-form Wilson loop symmetry at $h_x=0$, a string operator $\prod_{e\in L_{1/2}}Z_e$ creates a pair of charges at its two ends, where $L_{1/2}$ is an open string along the primal lattice. Thus, a string order parameter~\footnote{The string order parameter can also be interpreted as the ``disorder parameter" of the exact 1-form Wilson loop symmetry. In the deconfined topological phase, where the ground states spontaneously break the exact 1-form Wilson loop symmetry, the string order parameter vanishes. Conversely, the Higgs regime is disordered under the exact 1-form Wilson loop symmetry and the string order parameter is nonzero.}  $\tilde{O}_Z$ can be constructed to detect charge condensation~\cite{BRICMONT_1983}:
 \begin{equation}\label{order_para_h_z}
  \tilde{O}_Z=\lim_{|L_{1/2}|\rightarrow\infty}\sqrt{|\tilde{C}_Z(|L_{1/2}|)|}, \quad\quad \tilde{C}_Z(|L_{1/2}|)=\langle{\Psi}|\prod_{e\in L_{1/2}}Z_e|{\Psi}\rangle,
 \end{equation}
where $\ket{\Psi}$ is a normalized ground state of the toric code model, $|L_{1/2}|$ is the distance between two ends of $L_{1/2}$. The bulk of the string order parameter $\tilde{O}_Z$ commutes with the overlapping local Hamiltonian terms but not its ends. In the toric code phase, the string order operator thus creates two anyonic charge excitations that are orthogonal to the ground state, leading to a vanishing string order parameter in the infinite string limit, $\tilde{O}_Z=0$. By contrast, in the Higgs phase, the string order parameter can be nonzero because the $h_z$ field induces charge fluctuations such that charges condense in the Higgs phase. An alternative way of interpreting this string operator is by mapping $H_{\tTC}(0,h_z)$ to the $(2+1)$D transverse field Ising model~\cite{Wegner_duality_1971,Trebst_2007}. This transforms $\tilde{O}_Z$ and $\tilde{C}_Z$ to the Ising order parameter and its correlation function, respectively. From that also follows directly that the critical point $I$ [see Fig.~\ref{Figure_1}b] at $h_z=h_{zc}^{(I)}=0.328474(3)$~\cite{Youjin_2002} belongs to the (2+1)D Ising* universality class, where the ``*" indicates that in the effective Ginzburg-Landau-Wilson theory the order parameter field $\phi$ can only be created in pairs and $\phi$ and $-\phi$ are physically indistinguishable~\cite{torus_CFT_spec_2016}. 
Near the critical point $I$: $\tilde{O}_Z\sim(h_z-h_{zc}^{(I)})^\beta$ and $\xi\sim (h_z-h_{zc}^{(I)})^{-\nu}$, where  $\beta=0.326418(2)$ is the critical exponent of the order parameter~\cite{Conformal_bootstrap_2016}, $\xi$ is the correlation length defined via $|\tilde{C}_Z(|L_{1/2}|)-\tilde{O}_Z^2|\sim e^{-|L_{1/2}|/\xi}$, and $\nu=0.629 970(4)$ is the critical exponents of the correlation length~\cite{Conformal_bootstrap_2016}. 

\begin{table}[]
    \centering
    
    \begin{tabular}{cccccc}
    \toprule
           &   \multicolumn{2}{c}{$h_x=0$~\cite{BRICMONT_1983}} &  \multicolumn{2}{c}{$h_z=0$~\cite{Wegner_duality_1971}} & $h_x\& h_z\neq 0$~\cite{FM_1986}  \\ 
            \midrule
                &   deconf. & Higgs & deconf. & confined &   \\ 
           \midrule
        $\left\langle \prod_{e\in L}Z_e\right\rangle$ &  1&1   & $e^{-\alpha_Z|L|}$& $e^{-\alpha'_Z S_L}$  & $e^{-\alpha_Z|L|}$  \\  
       $\left\langle \prod_{e\in L_{1/2}}Z_e\right\rangle$ &  $e^{-\alpha'|L_{1/2}|}$&$O(1)$   & 0& 0  & $e^{-\alpha_Z|L_{1/2}|}$  \\ 
          \bottomrule
    \end{tabular}
    \caption{\textbf{$Z$ string and $Z$ loop operators.} The behavior of $Z$ loop (string) operator, defined as $\left\langle \prod_{e\in L}Z_e\right\rangle$ ($\left\langle \prod_{e\in L_{1/2}}Z_e\right\rangle$), are known from the references. When $h_z=0$, the ground states has the exact 1-form 't Hooft loop symmetry which anti-commutes $Z$-string operator, therefore $\left\langle \prod_{e\in L_{1/2}}Z_e\right\rangle=0$. $|L|$ ($|L_{1/2}|$) denotes the length of the loop (string) and $S_L$ is the size of the area surround by the loop $L$. $\alpha'$, $\alpha_Z$ and $\alpha'_Z$ are coefficients depending on details of models.}
    \label{tab:expec_loop_string}
\end{table}
 
When $h_x\neq0$, the Wilson loop operator $\prod_{e\in L}Z_e$ is no longer an exact 1-form symmetry of the toric code model and the bulk of the string $\prod_{e\in L_{1/2}}Z_e$ cannot deform freely. Therefore $\tilde C_Z(|L_{1/2}|)$ vanishes on either side of the topological phase transition exponentially with the length of the string $|L_{1/2}|$.  However, when $h_x$ is small, the toric code model has an emergent 1-form Wilson loop symmetry~\cite{Hastings_and_Wen_2005,Self_dual_critically_Ising,Wen_emergent_high_form_2023}. In the limit of large $h_x$, the 1-form Wilson loop symmetry cannot emerge, which we indicate by the gray and red areas in Fig.~\ref{Figure_1}b. In the presence of an emergent 1-form Wilson loop symmetry, one can in principle conceive to construct a dressed string operator 
with an extended width~\cite{cian2022extracting,cong_LED_2023}. This is however a challenging task in practice.
This problem can be circumvented, by dividing out the ``bulk'' contribution of the string order paramter, as proposed by Fredenhagen and Marcu, leading to the FM string order parameter~\cite{FM_1983,FM_1986}:
 \begin{equation}\label{FM_op}
   O_Z=\lim_{r\rightarrow\infty}\sqrt{|C_Z(r)|},\,\,\quad  C_Z(r)=\frac{\langle{\Psi}|\prod_{e\in L_{1/2}}Z_e|{\Psi}\rangle}{\sqrt{\bra{\Psi}\prod_{e\in L} Z_e\ket{\Psi}}},   
 \end{equation}
where $r=|L_{1/2}|$ is the length of the string $L_{1/2}$,\footnote{$|L_{1/2}|$ should be understood as the length of $L_{1/2}$ instead of the distance between its two ends because the bulk of the $Z$-string cannot be deformed freely when $h_x\neq 0$. In addition, for the $\mathbb{Z}_2$ gauge Higgs model, we should replace the string operator $\prod_{e\in L_{1/2}}Z_e$ with $Z_v\left(\prod_{e\in L_{1/2}}Z_e\right)Z_{v^{\prime}}$, where $v$ and $v^{\prime}$ are two vertices at the ends of the string $L_{1/2}$.} $L$ is a loop whose length is twice the length of the string $L_{1/2}$, see Fig.~\ref{Figure_1}a. Compared to Eq.~\eqref{order_para_h_z}, the FM string order parameter in Eq.~\eqref{FM_op} contains a square root of the expectation value of the Wilson loop operator in the denominator. When $h_x\neq 0$ the bare Wilson loop operator is no longer a symmetry and its expectation values decays in the presence of an emergent 1-form symmetry of charge excitations with a perimeter law (except in the flux condensation phase with exact 1-form 't Hooft loop symmetry ($h_z=0$), where the expectation value of the bare Wilson loop operator decays with an area law): 
$\bra{\Psi}\prod_{e\in L}Z_e\ket{\Psi}\sim\exp(-\alpha_Z|L|)$,
where $\alpha_Z$ is the perimeter law coefficient and $|L|$ is the length of the loop $L$. 
The denominator in the FM string order parameter  compensates the perimeter law decay of the numerator, such that only the contribution from the endpoints of the string $L_{1/2}$ is taken into account, where charges are created. We also separately summarize the behaviors of the numerator and denominator of the FM string order parameter in Table.~\ref{tab:expec_loop_string}, from which we immediately see that the FM string order parameter is always $0$ when $h_z=0$ and it cannot detect the phase transition between the deconfined phase and the confined phase. In addition, we emphasize that the FM string order parameter is a nonlinear function of the ground state. Therefore, one needs to carefully analyze whether it can serve as a \textit{bona fide} order parameter for topological phase transitions.
 
\begin{figure}
    \centering
    \includegraphics[scale=0.5]{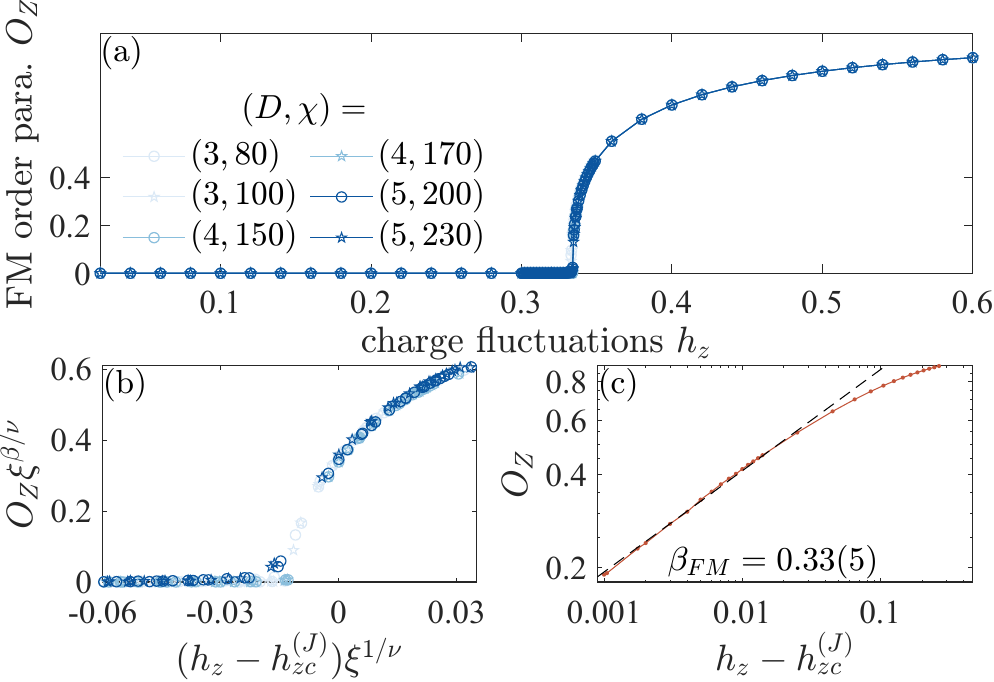}
    \caption{\textbf{FM string order parameter of the variational iPEPS along $h_x=0.3$.} 
    (a) The FM string order parameter from iPEPS with various bond dimensions $(D,\chi)$, see legend, is continuous across the topological phase transition. (b) Data collapse of the FM string order parameter with $\nu=0.629970(4)$ and $\beta=0.326418(2)$. (c) Double-log plot used for extracting the critical exponent $\beta_{\text{FM}}$ from the scaling of $O_Z$ in the vicinity of the critical point, where we we used $h_{zc}^{(J)}=0.335(1)$. Data is obtained by linearly extrapolating $O_Z$ in the inverse iPEPS bond dimension $1/D$. }
    \label{Figure_2}
\end{figure}

We now use iPEPS algorithms to directly evaluate the FM string order parameter in the limit of infinitely long strings on a torus, see Methods. When $L$ is a non-contractible loop on the torus, which is convenient for tensor network methods, care has to be taken when evaluating the denominator in Eq.~\eqref{FM_op}, as it could vanish for certain linear combinations of the degenerate ground states. Nonetheless, we are able to show that contractible and non-contractible loops are equivalent when the ground state is a minimally entangled state~\cite{MES_2012} with a trivial anyon flux penetrating through the torus; Fig.~\ref{Figure_1}a. Therefore, we use this choice for the ground state in the numerical evaluation of the FM order parameter; technical details are discussed in the Methods section.



\ \\ \ \\\textbf{FM string order parameter of the variational wave function.} 
We will now analyze the general properties of the FM string order parameter across a topological phase transition with charge condensation. In particular, we want to analyze whether the FM string order parameter is continuous upon crossing a second-order phase transition, and if so, we will analyze its scaling behavior. To this end, we first consider a cut through the phase diagram at $h_x=0.3$, that crosses the transition line $IM$ in Fig.~\ref{Figure_1}b at a point $J$. The criticality of the topological transition along the line $IM$ is expected to be described by the $(2+1)$D Ising* universality class. There are two important parameters that systematically control the error of the approximation when numerically evaluating the FM string order parameter: the bond dimension $D$ of the iPEPS itself and the bond dimension $\chi$ of the environment of the iPEPS; the larger bond dimensions provide better approximations. Therefore, we evaluate FM string string order parameter numerically along $h_x=0.3$ from iPEPS with various bond dimensions, see  Fig.~\ref{Figure_2}a. From this we find the critical point $J$ on the $h_x=0.3$ line is at $h_z=h_{zc}^{(J)}=0.335(1)$, consistence with $h_{zc}^{(J)}=0.333(1)$ from previous quantum Monte Carlo simulation~\cite{Youjin_2002}. In the toric code phase, the FM string order parameter vanishes while it is finite in the Higgs phase. 

\begin{table}[]
    \centering
    \begin{tabular}{cccccc}
    \toprule
          Exponent &   $(2+1)D$ &  $(2+0)D$   \\  
           \midrule
        $\Delta_{\phi}$ &  $0.5181489(10)$~\cite{Conformal_bootstrap_2016}   & $\frac{1}{8}$~\cite{CFT_book}     \\  
       $\Delta_{\phi^2}$ &  $1.412625(10)$~\cite{Conformal_bootstrap_2016}   &  $1$~\cite{CFT_book}   \\  
         $\nu = (\mathcal{D}-\Delta_{\phi^2})^{-1}$ & $0.629970(4)$ & $1$ \\
         \midrule
         $\beta = \Delta_{\phi} \nu$ & $0.326418(2)$ & $\frac{1}{8}$ \\
         $\beta_{\text{FM}}\; \mbox{(this work)}$ &  $0.33(5)$ & $0.117(5)$ \\ 
          \bottomrule
    \end{tabular}
    \caption{\textbf{Critical exponents of the Ising* field theory in $(2+1)$D and $(2+0)$D.} From the scaling dimensions $\Delta_\phi$ and $\Delta_{\phi^2}$ of the fields $\phi$ and $\phi^2$, respectively, one obtains the correlation length exponent $\nu$ and the critical exponent of the order parameter $\beta$, where $\mathcal{D}$ is the spacetime dimension. The critical exponent of the FM string order parameter $\beta_\text{FM}$ is obtained from iPEPS in this work. The critical exponents $\beta$ and $\beta_\text{FM}$ are consistent.}
    \label{tab:Ising}
\end{table}

Crucially, the FM string order parameter is continuous across a second order phase transition, such that we can extract the critical exponent $\beta_{\text{FM}}$ defined via $O_Z\sim(h_z-h^{(J)}_{zc})^{\beta_{\text{FM}}}$, Fig.~\ref{Figure_2}c, where we obtain $O_Z$ by performing linear extrapolation in $1/D$ and ignore the small $\chi$ dependence. The extracted critical exponents $\beta_{\text{FM}}=0.33(5)$ is consistent with the exponent $\beta=0.326418(2)$ of the $(2+1)$D Ising* universality class~\cite{Conformal_bootstrap_2016}, see Tab.~\ref{tab:Ising} for a summary of the exponents. We furthermore collapse the data from different bond dimensions, as shown in Fig.~\ref{Figure_2}b, based on the theory of finite entanglement scaling~\cite{Finite_Corboz_2018,Finite_Rader_2018,Bram_PEPS_scaling_2020}. The data for $D>2$ indeed collapses on a single curve near criticality. These results show that the FM string order parameter exhibits the correct critical behavior controlled by the $(2+1)$D Ising* universality class near the critical point $J$.

Previous order parameters for topological phase transitions were defined on the virtual legs of the iPEPS~\cite{Iqbal_2018,Iqbal_2017,Xu_2021,Xu_2022,Iqbal_Schuch_2021}, because there exist the virtual symmetries in terms of matrix product operators~\cite{Schuch_2010,Anyon_MPO_2017,characterizing_MPO_2021}. However, without the iPEPS representation, these virtual order parameters cannot be obtained, i.e., they are not physical. This should be contrasted with the FM string order parameter that is defined on the physical level and thus also does not depend on the iPEPS gauge. %

\begin{figure*}
    \centering
    \includegraphics[scale=0.5]{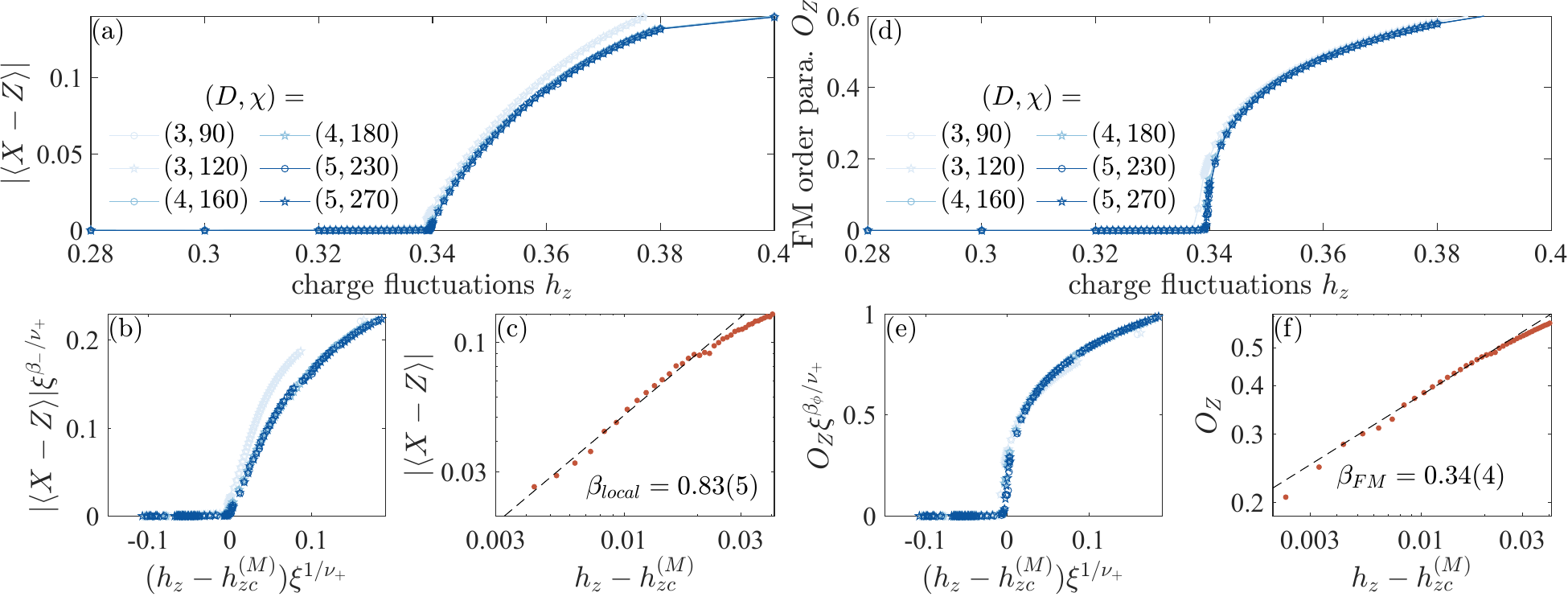}
    \caption{\textbf{Local duality symmetry-breaking and FM string order parameters of the variational iPEPS along the self-dual line $h_x=h_z$.} (a) Local order parameter $|\langle X-Z \rangle|$ from iPEPS with various bond dimensions $(D,\chi)$, see legend. (b) Data collapse of the local order parameter with $\nu_+=0.67175(1)$ and $\beta_-=0.83048(2)$. (c) Critical exponent $\beta_{\text{local}}$, where $|\langle X-Z\rangle|$ is obtained by linear extrapolation in $1/D$ and $h^{(M)}_{zc}=0.3397(2)$. (d) FM string order parameter from iPEPS with various bond dimensions, see legend. (e) Data collapse of the FM string order parameter with $\beta_{\phi}=0.34870(7)$. (f) Critical exponent $\beta_{\text{FM}}$, where $O_Z$ is again obtained by linear extrapolation in $1/D$.}
    \label{Figure_3}
\end{figure*}

Next, we consider the FM string order parameter  of the variational wave function  along the self-dual line {$h_x=h_z$}, where the toric code model in Eq.~\eqref{TC_Hamiltonian} has a global electric-magnetic duality symmetry, which exchanges the primal lattice and the dual lattice, as well as $X$ and $Z$. There is a gapped electric-magnetic duality symmetry breaking phase $MK$ along the self-dual line~\cite{TC_phase_diagram_expansion_2009,Youjin_TC_phase_diagram_2011,Self_dual_critically_Ising}. The transition between the toric code phase and the duality symmetry breaking phase is a multi-critical point $M$ shown in Fig.~\ref{Figure_1}b. From our iPEPS simulation we find that  $M$ is located at $h_x=h_z=h^{(M)}_{zc}=0.3397(2)$, which is again close to $0.340(2)$ from the quantum Monte Carlo simulations~\cite{Youjin_TC_phase_diagram_2011} and to $0.3406(4)$ obtained from the higher-order perturbation expansion~\cite{TC_phase_diagram_expansion_2009}. The phase transition crossing the multi-critical point $M$ can be characterized by a local symmetry-breaking order parameter $|\langle X-Z\rangle|$, which is the difference between expectation values of of $X_e$ and $Z_e$ on a single edge $e$, as shown in Fig.~\ref{Figure_3}a, from which we extract a critical exponent  $\beta_{\text{local}}=0.83(5)$ defined via $|\langle X-Z\rangle|\sim (h_z-h_{zc}^{(M)})^{\beta_{\text{local}}}$,  see Fig.~\ref{Figure_3}c. Our result is consistent with recent Monte Carlo simulation~\cite{Self_dual_critically_Ising}. 
We also evaluate the FM string order parameter along the self-dual line crossing the multi-critical point $M$; Fig.~\ref{Figure_3}d. For the FM string order parameter, we extract the critical exponent $\beta_{\text{FM}}=0.34(4)$ defined via $O_Z\sim(h_z-h^{(M)}_{zc})^{\beta_{\text{FM}}}$, Fig.~\ref{Figure_3}f, which is quite different from the critical exponent of the local symmetry-breaking order parameter $\beta_{\text{local}}=0.83(5)$. 

How can we understand the distinct critical exponents $\beta_{\text{local}}$ and $\beta_{\text{FM}}$? Since the phase boundary $IM$ belongs to the Ising* universality class and the FM string order parameter exhibits an Ising critical exponent, we can assume that the FM string order parameter $O_Z$ corresponds to a field $\phi_z$ of an effective Ginzburg-Landau-Wilson theory describing the Ising* transition. When two Ising* transition lines $IM$ and $GM$ in Fig.~\ref{Figure_1}b meet at the multi-critical point $M$, once could conceive that the effective field theory of the multi-critical point $M$ is the XY* model with a Lagrangian $\mathcal{L}=\left(\partial \pmb{\phi}\right)^2/2+m^2\pmb{\phi}^2/2+g\pmb{\phi}^4/(4!)$~\cite{Self_dual_critically_Ising,Multicritical_2022}, up to some irrelevant terms, which possesses $O(2)$ symmetry. Here, $\pmb{\phi}=(\phi_x,\phi_z)$ is a two-component vector field. We summarize the scaling dimensions of the order parameter fields and relevant critical exponents in Tab.~\ref{tab:O2}. We find that $\beta_\text{FM}$ is consistent with $\beta_\phi$ and $\beta_\text{local}$  with $\beta_-$ of the XY* field theory.  This can be understood by identifying $\phi_z$ with the FM string order parameter $O_z$, and $\phi_x^2-\phi_z^2$ with $X-Z$~\cite{Self_dual_critically_Ising,Multicritical_2022,Machine_learning_TC_SD_2023}.    
We can separately check consistency with the  XY* field theory by performing data collapse of the FM string order parameter and the local order parameter using the critical exponents of the field theory; Figs.~\ref{Figure_3}b and e, where we find a reasonable collapse for iPEPS dimension $D>3$.

\begin{table}[]
    \centering

  \begin{tabular}{cccccc}
    \toprule
          Exponent &   $(2+1)$D &  $(2+0)$D   \\  
           \midrule
        $\Delta_{\phi}$ &  $0.519088(22)$~\cite{O_2_exponent_2020} & $\frac{1}{8}$~\cite{Nienhuis_1982}     \\  
       $\Delta_{+}$ &  $1.51136(22)$~\cite{O_2_exponent_2020}  &  $1$~\cite{Nienhuis_1982}  \\  
        $\Delta_{-}$ &  $1.23629(11)$~\cite{O_2_exponent_2020}  &  $\frac{1}{2}$~\cite{Coulomb_gas_1987}   \\  
         $\nu_{+}=(\mathcal{D}-\Delta_{+})^{-1}$ & $0.67175(1)$ & $1$ \\
         \midrule
         $\beta_{\phi}=\Delta_{\phi} \nu_{+}$ & $0.34870(7)$ & $\frac{1}{8}$   \\ 
          $\beta_{-}=\Delta_{-} \nu_{+}$ & $0.83048(2)$ &  $\frac{1}{2}$\\
           $\beta_{\text{FM}} \; \mbox{(this work)}$ & $0.34(4)$ & $-$ \\ 
           $\beta_{\text{local}}  \; \mbox{(this work)}$ & $0.83(5)$ & $-$ \\      
          \bottomrule
\end{tabular}

    \caption{\textbf{Critical exponents of the XY* field theory in (2+1)D and (2+0)D.} Scaling dimensions $\Delta_\phi$, $\Delta_+$, and $\Delta_-$ of the fields $\phi_z$, $\pmb{\phi}^2 = \phi^2_x+\phi^2_z$, $\phi^2_x-\phi^2_z$, respectively. From these scaling dimensions, one obtains the critical exponent $\nu_{+}$ of the correlation length, $\beta_\phi$ of the single component of the order parameter $\phi_x$,  and $\beta_-$ of the additional order parameter $\phi^2_x-\phi^2_z$. $\mathcal{D}$ is the spacetime dimension. The critical exponent of the FM string order parameter $\beta_\text{FM}$ and the local order parameter $\beta_\text{local}$ obtained from iPEPS in this work are consistent with $\beta_\phi$ and $\beta_-$ in $(2+1)$D, respectively. In $(2+0)$D, since there is a gapless BKT phase along the self-dual line in Fig.~\ref{FM_op_deform_TC}a, we can not define $\beta_{\text{FM}}$ and $\beta_{\text{local}}$. 
    }
    \label{tab:O2}
\end{table}

\begin{figure*}
    \centering
    \includegraphics[scale=0.5]{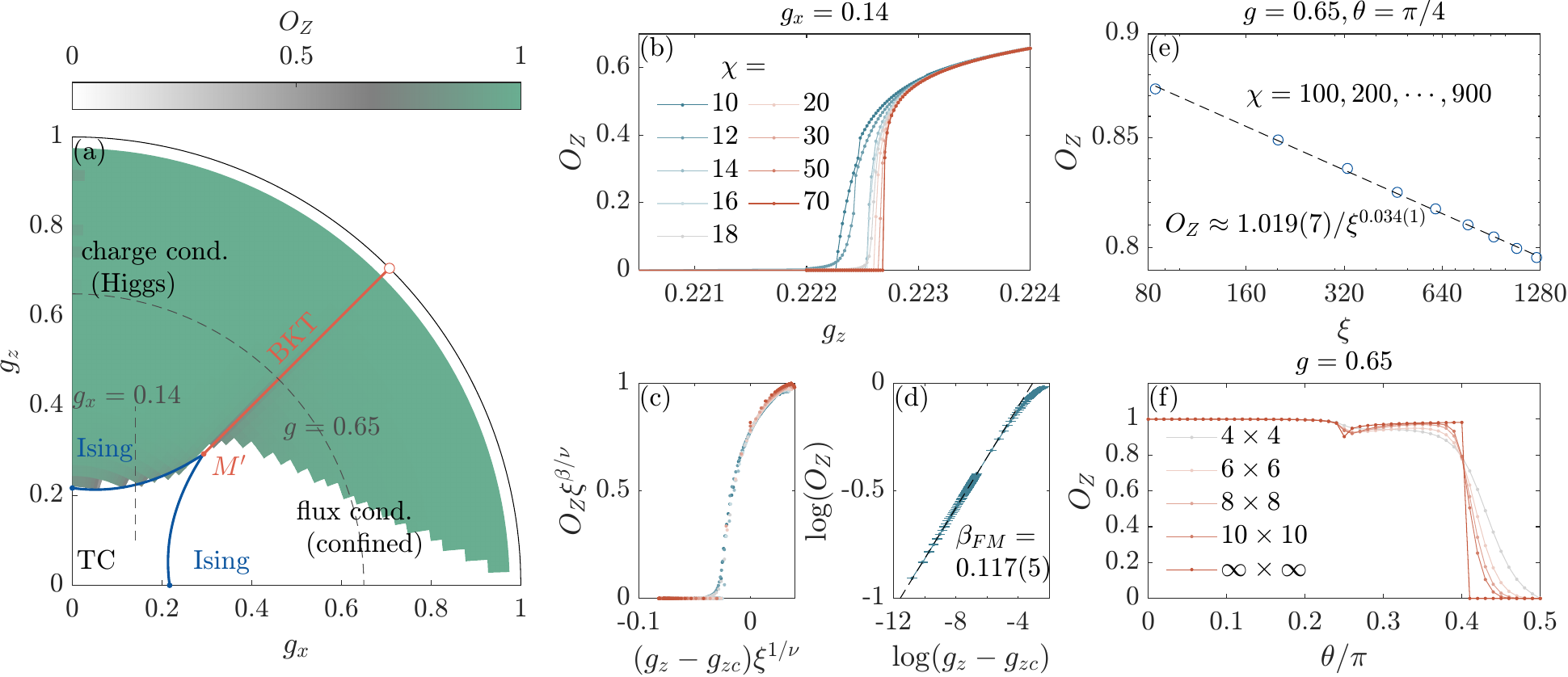}
    \caption{\textbf{FM string order parameter for the deformed toric code wave function.} (a) Phase diagram of the deformed toric code state of Eq.~\eqref{deformed_wavefunction}. We also plot a color-map of the FM order parameter $O_Z$ in the phase diagram. (b) The FM string order parameter along $g_x=0.14$ [vertical dashed line in (a)], calculated using iMPS with various bond dimensions $\chi$. (c) Data collapse of the FM string order parameter with $g_{zc}=0.2227(3)$, $\beta=1/8$ and $\nu=1$. (d) Critical exponent $\beta_{\text{FM}}$ of the FM string order parameter, obtained by extrapolating $O_Z$ for different $\chi$. (e) Double-log plot of the FM string order parameter $O_Z$ as a function of the correlation length $\xi$ of the boundary MPS at the point $(g,\theta)=(0.65,\pi/4)$ in the BKT phase. $O_z$ vanishes in the limit of infinite bond dimension $\chi\to \infty$. (f) Finite and infinite length FM string order parameters along $g_x^2+g_z^2=0.65^2$ [dashed quarter circle in (a) parametrized with $\theta=\arctan(g_x/g_z)$] from the iMPS with $\chi=20$. The legend shows the size of the area surrounded by the loop operator.}
    \label{FM_op_deform_TC}
\end{figure*}

Some comments are in order. First, the two degenerate ground states in the duality symmetry-breaking phase correspond to the predominant condensation of charges (fluxes) satisfying $\langle X-Z\rangle<0$ ($\langle X-Z\rangle>0$). Because our FM string order parameter is defined with a $Z$ string, it detects the charge condensation, and thus, we should use the charge condensation-dominated state to evaluate the FM string order parameter. In contrast, the dual FM string order parameter associated with the flux excitations is not well-behaved when applied to the same ground state; see supplemental materials~\cite{appendix}. 
Second, Ref.~\cite{Self_dual_critically_Ising} argues that the multi-critical point $M$ may not belong to the XY* universality class because the mutual semionic statistics between charges and fluxes is not included in the  XY* field theory in an obvious way.  
Although our numerical data is consistent with the XY* field theory, we cannot rule out other field theoretic descriptions of the transition due to the comparatively large uncertainties in the critical exponents. 

\ \\\textbf{FM string order parameter for the deformed toric code state.} 
Instead of variationally solving the ground state Hamiltonian in Eq.~\eqref{TC_Hamiltonian}, one can analytically construct a deformed toric code state, which shares similar physics with the toric code Hamiltonian in Eq.~\eqref{TC_Hamiltonian}. The advantage of the deformed toric code state is that the wavefunction is exact, so many analytical results can be derived. The disadvantage is that the dimensionality of the universality class of the quantum critical points is fine-tuned and reduced by 1 compared to that of the generic quantum critical points~\cite{Simons_Trebst_2009,Near_and_at_CQCP_2011}. The deformed toric code state is defined as~\cite{haegeman2015shadows,Gauging_quantum_state_2015,Zhu_2019}: 
\begin{equation}\label{deformed_wavefunction}
    \ket{\psi(g_x,g_z)}=\prod_e(1+g_x X_e+g_z Z_e)\ket{\tTC},
\end{equation}
where $\ket{\tTC}$ is a ground state of the fixed point toric code Hamiltonian, and $g_x$ and $g_z$ are tuning parameters satisfying $g_x^2+g_z^2\leq1$. The phase diagram of the deformed toric code state is similar to that of the toric code Hamiltonian in Eq.~\eqref{TC_Hamiltonian}, as shown in Fig.~\ref{FM_op_deform_TC}a. Besides the toric code phase, there are two trivial phases with either charge or flux condensation, and they are separated by a Berezinskii-Kosterlitz-Thouless (BKT) transition line. The phase transition lines from the toric code phase to the trivial phase belong to the (2+0)D Ising* universality class because of the dimensionality reduction at the fine-tuned critical points of the wave function. Because the deformed toric code state is mapped to the partition function of the 2D classical Ashkin-Teller model~\cite{Zhu_2019}, the multi-critical point $M'=(1-1/\sqrt{2},1-1/\sqrt{2})$ at the self-dual line $g_x=g_z$ is described by the free boson conformal field theory (CFT) compactified on orbifolds with a radius $R=2\sqrt{2}$, which describes the critical point of the 2D XY* theory~\cite{O_N_CFT_2023}. Thus, we can say the multicritical point $M'$ of the deformed toric code state belongs to the (2+0)D XY* universality class. 

Akin to the variational case, we first calculate the FM string order parameter along a line $g_x=0.14$ crossing the toric code phase and the Higgs phase by contracting the $D=2$ exact iPEPS of the deformed toric code state, using the boundary infinite matrix product states (iMPS) with various bond dimensions $\chi$,  
see Fig.~\ref{FM_op_deform_TC}b. We extract the critical exponent $\beta_{\text{FM}}$ of the FM string order parameter according to $O_Z\sim (g_z-g_{zc})^{\beta_{\text{FM}}}$, for which we numerically determine the critical point $g_{zc}=0.2227(3)$ and extrapolate $O_Z$ using different iMPS bond dimensions $\chi$. The extracted $\beta_{\text{FM}}=0.117(5)$ is close to the $\beta=1/8=0.125$ from the 2D Ising* universality class, see  Fig.~\ref{FM_op_deform_TC}d and Tab.~\ref{tab:Ising}. We perform the data collapse 
to the FM string order parameter in Fig.~\ref{FM_op_deform_TC}c. These results indicate that the FM string order parameter is a well-behaved order parameter also for the deformed wavefunction. {In contrast to the variatinoal case, along the self-dual line the phase transition from the toric code phase to the BKT phase along the self-dual line is not a charge condensation transition, and the FM string order parameter is zero in both the toric code phase and the BKT phase, see Fig.~\ref{FM_op_deform_TC}e. Therefore,  the FM string order parameter cannot be directly used to detect the BKT transition.}



The FM string order parameter evaluation of the deformed toric code is numerically stable, since the wave function can be exactly expressed in terms of an iPEPS without the need of variational optimization. We evaluate the FM string order parameter in the entire phase diagram of the deformed toric code state, as shown in Fig.~\ref{FM_op_deform_TC}a as a color plot,  which surprisingly exhibits a sharp transition in the flux condensation phase even in the absence of a quantum phase transition. To emphasize this behavior, we  show the FM string order parameter along a path $g^2_x+g^2_z=0.65^2$ in Fig.~\ref{FM_op_deform_TC}f. It is discontinuous at a certain angle $\theta=\arctan(g_x/g_z)\approx 0.4\pi$ even in the absence of a bulk phase transition. In order to exclude the possibility that there are some artifacts of our method that cause the discontinuity, we also evaluate the FM string order parameter with finite string length $r=|L_{1/2}|$ in Fig.~\ref{FM_op_deform_TC}f. As the string length increases, the results from the thermodynamic limit are obtained, which indeed implies that the FM string order parameter becomes discontinuous for infinite string lengths. 
{Moreover, there is a small drop at the self-dual line in Fig.~\ref{deformed_wavefunction}f, which slowly approaches $0$ with increasing bond dimension; Fig.~\ref{deformed_wavefunction}e.}

Although surprising, the discontinuity in the flux condensed phase is possible because the FM string order parameter is a non-linear function of the ground state. The underlying reason for this abrupt change of the FM order parameter is that the parity of the dominant eigenvectors of the transfer matrices, whose overlap determines the FM order parameter, change at the discontinuity~\cite{appendix}. In the supplemental materials~\cite{appendix}, we show that for the Hamiltonian in Eq.~\eqref{TC_Hamiltonian} a similar singular behavior of the FM string order parameter can be found when calculating the FM order parameter using an iPEPS constructed via perturbing the infinite-field limit product states in the confined region~\cite{Laurens_bridge_2017}.
These unexpected results indicate that in the flux condensation region, the FM string order parameter that creates charges at its end cannot be used as an order parameter anymore, and the FM string order parameter does not have a physical significance. We discuss the consequences of this behavior in the next section.

\ \\{\large\textbf{Discussion}}\\
We have evaluated the FM string order parameter in the infinitely long string limit using the iPEPS simulation and found that it exhibits universal scaling controlled by the underlying critical points of charge condensation transitions. 

Our results indicate that the FM string order parameter can be discontinuous in the flux condensation region, which does not possess an emergent 1-form Wilson loop symmetry, see Figs.~\ref{FM_op_deform_TC}a and f. Hence, we argue that only in the presence of an emergent 1-from Wilson loop symmetry, the associated FM string order parameter can be a quantitatively well-behaved order parameter for topological phase transitions. 
Our argument can be formulated using the idea of quantum error correction~\cite{Sagar_ViJay_2024,Adam_Nahum_2024}. A quantum state $\ket{\Psi}$ having an emergent 1-form Wilson loop symmetry implies that we can apply a recovery map constructed from quantum error correction to $\ket{\Psi}$ to get an exact 1-form Wilson loop symmetric state $\ket{\Psi_0}$ within the same phase of $\ket{\Psi}$. Applying this recovery map to both the numerator and denominator of the FM order parameter, we have $\bra{\Psi}\prod_{e\in L_{1/2}} Z_e\ket{\Psi}\sim e^{-\alpha |L_{1/2}|}\bra{\Psi_0} \prod_{e\in L_{1/2}} Z_e\ket{\Psi_0}$ and  $\bra{\Psi}\prod_{e\in L_{1/2}} Z_e\ket{\Psi}\sim e^{-\alpha |L|}
$~\cite{QEC_and_1_form}, where $\alpha$ is a decay coefficient that depends on $\ket{\Psi}$ and the recovery map. The FM order parameter evaluated with $\ket{\Psi}$ using Eq.~\eqref{FM_op} then reduces to the string order parameter evaluated with $\ket{\Psi_0}$ using Eq.~\eqref{order_para_h_z}. As a consequence, the FM order parameter is a well-behaved order parameter in the presence of emergent 1-form Wilson loop symmetry. By contrast, if there is no emergent 1-form Wilson loop symmetry, it will not be guaranteed that a recovery map exists. 
In the confined regime, the magnetic 1-form 't Hooft loop symmetry is emergent, and one has to construct an FM order parameter of $X$ operators along the dual lattice to detect the phase transition.
Therefore, when detecting a topological phase transition using FM string order parameters, either with numerical simulations or experiments, knowledge about the underlying emergent 1-from symmetries is required.   

\begin{figure*}
    \centering
    \includegraphics{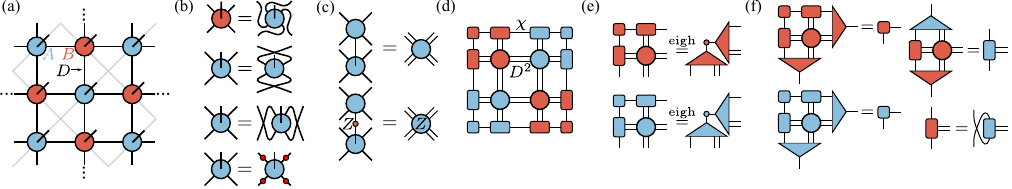}
    \caption{\textbf{iPEPS ansatz and the CTMRG.} (a) The iPEPS ansatz for the ground state of the toric code model with a bond dimension $D$. (b) The iPEPS tensors $A$ and $B$ are related by rotation. The iPEPS tensor $A$ is invariant under two reflections, and we can also impose the virtual $\mathbb{Z}_2$ symmetry when necessary, where the red dots are matrices $Z_D$. (c) A double tensor and another double tensor sandwiching a $Z$ matrix. (d) The environment of the iPEPS is approximated by the corner tensors (squares) and the edge tensors (rectangles). (e) The hermitian eigenvalue decomposition (eigh) of the top-left corner of the tensor networks in (d), the isometries can be obtained from eigenvectors corresponding to the $\chi$ largest eigenvalues (in absolute value). (f) CTMRG procedures updating the corner and edge tensors using the isometries.}
    \label{Fig_append_1}
\end{figure*}

Our work opens several questions and research directions. First, measuring the FM string order parameter in quantum simulation experiments can provide insights into topological phase transitions with charge condensation. 
Second, in Ref.~\cite{QEC_and_1_form}, we argue that 1-form symmetries can be detected with quantum error correction. It will be interesting to precisely analyze the boundaries of the three regions of different emergent 1-form symmetries in the trivial phase of the toric code model in Eq.~\eqref{TC_Hamiltonian}, shown schematically in Fig.~\ref{Figure_1}b, and investigate the nature of the transitions between presence and absence of the 1-form symmetries. 
Third, the FM string order parameter can be applied to different lattice gauge theories~\cite{FM_OP_2011,Zohar_Nussinov_2013,Ke_Liu_2015,BEEKMAN20171}. It is an interesting direction to define the FM string order parameters for Kitaev's quantum double models~\cite{kitaev_2002} and Levin-Wen string-net models~\cite{String_net_2005} and apply them to study various topological phase transitions driven by anyon condensation~\cite{Xu_2020,Xu_2021,Xu_2022}. Moreover, since the Higgs phase can be interpreted as a phase protected simultaneously by both a 1-form symmetry and a global symmetry~\cite{Ruben_Higgs_SPT}, one could use the FM-type string order parameters to detect such kind of symmetry-protected topological order, especially when the protecting 1-form symmetry becomes an emergent symmetry.

\ \\ \ \\
{\large\textbf{Methods}}\\
\textbf{iPEPS optimization.}
In this section, we show the technical details of optimizing the ground states of the toric code model using iPEPS. 
We approximate a ground state using the iPEPS ansatz proposed in Ref.~\cite{Corboz_2020}. The iPEPS has a $2\times2$ unit cell, and it is parameterized by a rank-5 tensor $A$ ($B$ is obtained from $A$ by a $\pi/2$ rotation) with the virtual bond dimension $D$ and the physical dimension $d=2$, as shown in Figs.~\ref{Fig_append_1}a and b. 

We impose the square lattice symmetry onto the tensor $A$ such that the iPEPS tensor is invariant under two reflections $R_v$ and $R_h$, see Fig.~\ref{Fig_append_1}b. Because of the symmetry, the number of independent variational parameters is less than $2D^4$. We can parameterize such a symmetric tensor $A$ using the following method. We first construct the $2D^4\times 2D^4$ matrix representations of $R_h$ and $R_v$ applying on the tensor $A$, and a projector $P_R=(\mathbbm{1}+R_h)(\mathbbm{1}+R_v)/4$. 
A subspace spanned by the eigenvectors of $P_R$ with an eigenvalue 1 is $\{\ket{v_i}|P_R\ket{v_i}=\ket{v_i}\}$. If $\braket{v_i}{v_j}\neq\delta_{ij}$, we can orthonormalize them using the QR decomposition. Reshaping $\ket{v_i}$ to the tensors with the dimensions $D\times D\times D\times D\times 2$, we can parameterize the iPEPS tensor as $A=\sum_i\lambda_iv_i$, where $\lambda=\{\lambda_i\}$ are variational parameters. If we also want to impose the virtual $\mathbb{Z}_2$ symmetry to the tensor $A$ as shown in Fig.~\ref{Fig_append_1}a, we need another projector $P_Z=(\mathbbm{1}_D^{\otimes 4}\otimes \mathbbm{1}_d+Z_D^{\otimes 4}\otimes \mathbbm{1}_2)/2$, where $Z_D$ is a $D\times D$ matrix representation of the non-trivial element in $\mathbb{Z}_2$, i.e., $Z_D^2=1$; and we consider the projector $P=P_ZP_R$, where $[P_Z,P_R]=0$. Using the orthonormal basis of the subspace spanned by the eigenvectors of $P$ with eigenvalue $1$, we can parameterize the tensor $A$ using $\lambda$. 

The iPEPS $\ket{\Psi(\lambda)}$ can be constructed from $A(\lambda)$. The energy expectation value $E=\bra{\Psi(\lambda)}H_{\text{TC}}(h_z,h_x)\ket{\Psi(\lambda)}/\braket{\Psi(\lambda)}{\Psi(\lambda)}$ can be evaluated by contracting iPEPS $\ket{\Psi(\lambda)}$. 
We contract (the squared norm of) the iPEPS using the corner transfer matrix renormalization group (CTMRG) algorithm. As shown in Fig.~\ref{Fig_append_1}d, we approximate the environment of the double tensors (shown in Fig.~\ref{Fig_append_1}c) in a $2\times2$ unit cell using corner (rectangles) and edge tensors (squares)  with a bond dimension $\chi$. Since we impose the square lattice symmetry to the tensor $A$, we can contract the iPEPS using the symmetric CTMRG~\cite{Corboz_2020}. 
The bond dimensions of the corner and edge tensors grow to $D^2\chi$ after absorbing the double tensors, and we should truncate the bond dimension back to $\chi$.  Fig.~\ref{Fig_append_1}e shows that using hermitian eigenvalue decomposition we obtain the isometries (triangles), which can be used to truncate the bond dimensions,  and we just use eigenvectors corresponding to the $\chi$ largest eigenvalues (in absolute value) to construct the isometry. With the isometries, we can update corner and edge tensors, as shown in Fig.~\ref{Fig_append_1}f. 

In order to optimize the iPEPS, one has to provide the energy gradient $\partial E/\partial \lambda$. The best way to calculate the energy gradient is using automatic differentiation (AD)~\cite{Liao_AD_2019}, which calculates the gradient through a backward propagation along the computational graph based on the chain rule in calculus.  One problem of applying AD to calculate $\partial E/\partial \lambda$ is that the gradient can be infinite when eigenvalues are degenerate. Although one can add a small perturbation to lift the degeneracy, numerical instability can still happen with a small probability. When we get an infinity gradient, we can detach the isometries from the computation graph and get an approximate gradient; a trade-off between the stability and the accuracy. A possibly better solution is to use the approaches shown in Ref.~\cite{Anna_2023}.   

 Given the energy expectation value and its gradient, we use the BFGS (Broyden–Fletcher–Goldfarb–Shanno) algorithm to minimize the energy expectation value. When the optimization is converged, we have an iPEPS $\ket{\Psi(\lambda)}$ approximating a ground state of the toric code model. For this work, the iPEPS optimization was performed by PyTorch on the NVIDIA A100 80 GB GPU cards. We use the checkpoint function of PyTorch to reduce the huge memory cost of backward AD calculation. Moreover, when getting the isometries, we should use the hermitian eigenvalue decomposition rather than singular eigenvalue decomposition, because the former is about ten times faster than the latter on the GPU. It takes about two weeks (3 days) to calculate a single curve containing more than 100 data points  with bond dimension $D=5$ ($D=4$) on a single A100 GPU card. Near critical points, CTMRG needs about 200 to 300 iterations to converge for a single energy evaluation.

\ \\ \ \\ \\\textbf{iPEPS for topologically degenerate ground states.} 
Here, we discuss two kinds of iPEPS representations of the toric code ground states. Understanding them is useful for initializing the iPEPS optimization and evaluating the FM string order parameter. When $h_x=h_z=0$, the toric code Hamiltonian defined on a torus commutes with two Wilson loop operators and two 't Hooft loop operators:
\begin{equation}\label{four_wilson_loop}
W^{Z}_{x}=\prod_{e\in L_x } Z_e, \quad W^{Z}_{y}=\prod_{e\in L_y } Z_e,\quad W^{X}_{x}=\prod_{e\in \hat{L}_x } X_e, \quad W^{X}_{y}=\prod_{e\in \hat{L}_y } X_e,
\end{equation}
where $L_x$ ($L_y$) is a non-contractible loop along $x$($y$) direction on the primal lattice, and $\hat{L}_x$ ($\hat{L}_y$) is a non-contractible loop along $x$($y$) direction on the dual lattice. They satisfy
\begin{align}
\left[W_x^Z,W_y^Z\right]&=\left[W_x^X,W_y^X\right]=\left[W_x^Z,W_x^X\right]=\left[W_y^Z,W_y^X\right]=0,\notag\\
\left\{W_x^Z,W_y^X\right\}&=\left\{W_x^X,W_y^Z\right\}=0.
\end{align}
The four ground states can be labeled by eigenvalues of a pair of two commuting loop operators, i.e., common eigenstates of $W^{X}_{x}$ and $W^{X}_y$:
\begin{equation}\label{GS_XX_basis}
    \ket{+_x+_y},\quad\ket{+_x-_y},\quad\ket{-_x+_y},\quad \ket{-_x-_y},
\end{equation}
or common eigenstates of $W^{Z}_{x}$ and $W^{Z}_y$:
\begin{equation}\label{GS_ZZ_basis}
\ket{0_x0_y},\quad\ket{0_x1_y},\quad\ket{1_x0_y},\quad\ket{1_x1_y}.
\end{equation}
In particular, the common eigenstates of $W_x^Z$ and $W_x^X$ (alternatively one can use 
$W_y^Z$ and $W_y^X$) are minimally entangled states~\cite{MES_2012}: 
\begin{equation}\label{GS_MES}
\ket{0_x+_x}\equiv\ket{\pmb{1}},\quad\ket{1_x+_x}\equiv\ket{\pmb{m}},\quad\ket{0_x-_x}\equiv\ket{\pmb{e}},\quad\ket{1_x-_x}\equiv\ket{\pmb{f}}.
\end{equation}

\begin{figure*}
    \centering
    \includegraphics{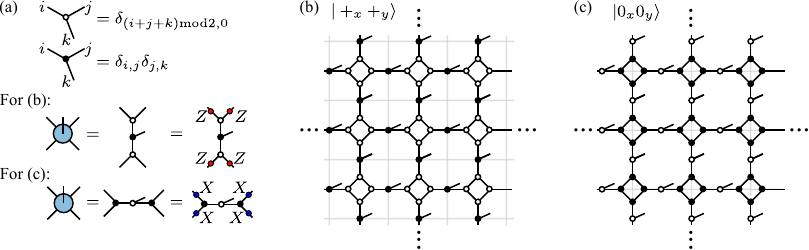}
    \caption{\textbf{The exact iPEPS at the fixed point $h_x=h_z=0$ of the toric code phase.} (a) The two rank-3 three tensors are defined to construct the tensor $A$. The tensor $A$ for $\ket{+_x+_y}$ and $\ket{0_x0_y}$ has different virtual $\mathbb{Z}_2$ symmetry. iPEPS for (b) $\ket{+_x+_y}$ and (c) $\ket{0_x0_y}$ on a torus. The gray lines indicate the primal lattice.}
    \label{Fix_point_PEPS}
\end{figure*}

 Let us focus on the ground states $\ket{0_x0_y}$ and $\ket{+_x+_y}$, which can be exactly expressed in terms of the so-called ``single-line" and ``double-line" iPEPS with the toroidal boundary condition~\cite{Single_line_TNS_2008}. They are included in the $2\times2$ unit cell iPEPS ansatz in Fig.~\ref{Fig_append_1}a. As shown in  Fig.~\ref{Fix_point_PEPS}a, by defining two of rank-3 tensors, we can obtain the $A$ tensor for the iPEPS $\ket{+_x+_y}$ shown in Fig.~\ref{Fix_point_PEPS}b or the iPEPS $\ket{0_x0_y}$ shown in Fig~\ref{Fix_point_PEPS}c. Because the two kinds of nonequivalent of iPEPS representations of the fixed point toric code states are included in the  $2\times 2$ unit cell iPEPS ansatz, we expect that it performs better than other kinds of iPEPS ansatz for the toric code model, especially along the self-dual line.

\begin{figure}
    \centering
    \includegraphics{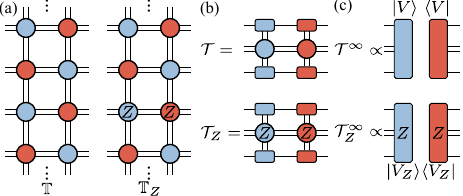}
    \caption{\textbf{Evaluating the FM string order parameter using iPEPS.}  (a) Using the double tensors in Fig.~\ref{Fig_append_1}c, two iPEPS transfer matrices are defined to evaluate the FM string order parameter. (b) Compressing the iPEPS transfer matrices using the edge tensors from the CTMRG. (c) The infinite power transfer matrices are given by their fixed points.}
    \label{Figure_1_prime}
\end{figure}

When evaluating the FM string order parameter on a torus geometry for a non-contractible loop $L$, which as we have seen is convenient for our tensor network methods, the denominator in Eq.~\eqref{FM_op} depends on the choice of the topologically degenerate ground states in the toric code phase.
This can be easily understood at the fixed point of the toric code phase. When using the ground state $\ket{+_x+_y}$ to evaluate the FM string order parameter $O_Z$, we find that it is not well-defined because $\ket{+_x+_y}$ spontaneously breaks the $W^Z_x$ symmetry, i.e., $W^Z_x\ket{+_x+_y}=\ket{+_x-_y}$, such that the denominator $\bra{+_x+_y}W^Z_x\ket{+_x+_y}=0$. This problem can be circumvented by considering either the ground state $\ket{0_x0_y}$ or the minimally entangled state of the trivial topological sector $\ket{\pmb{1}}$, 
such that the denominator of the FM string order parameter $O_Z$ is $1$. However, if we want to correctly evaluated FM string order parameters $O_X$ and $O_Z$ simultaneously, we have to consider the minimally entangled state $\ket{\pmb{1}}$. The FM string order parameters $O_X$ and $O_Z$ evaluated using the minimally entangled state in the trivial topological sector which mimics the one defined on a contractible loop. 

Away from the fixed point of the toric code model, the exact Wilson and 't Hooft loop operators in Eq.~\eqref{four_wilson_loop} are not symmetries the Hamiltonian anymore. Since we mainly focus on the FM string order parameter $O_Z$, we just need to find the ground state $\ket{0_x0_y}$, which is the simultaneous eigenstate of emergent non-contractible Wilson loop operators along $x$ and $y$ directions. 
Because the optimized iPEPS usually converge to $\ket{0_x0_y}$ or $\ket{+_x+_y}$, we can control which ground state it converges to by initializing the iPEPS tensor as $A+\epsilon R$, where $A$ is given in Fig.~\ref{Fix_point_PEPS}a and $R$ is a random tensor satisfying the required lattice symmetry, and $\epsilon$ is a small number. We emphasis that we do not imposing the virtual $\mathbb{Z}_2$ symmetry shown in Fig.~\ref{Fig_append_1}b to the tensor during the optimization. 

\ \\ \ \\ \\\textbf{Evaluation of the FM string order parameter using iPEPS.} With the optimized iPEPS tensor, we can evaluate the FM string order parameter efficiently in the limit of an infinitely long string using transfer matrices of iPEPS. To this end, we construct from the iPEPS of the ground state two transfer matrices; one is the usual transfer matrix $\mathbb{T}$ and the other is  $\mathbb{T}_Z$ containing $Z$ operators; see Fig.~\ref{Figure_1_prime}a for graphical notations. The numerator of the FM string order parameter consists of $r$ transfer matrices $\mathbb{T}$ followed by $r$ transfer matrices $\mathbb{T}_Z$ containing a $Z$ string and the denominator consists of $2r$ transfer matrices $\mathbb{T}_Z$ containing a $Z$ loop:
\begin{align}
    O_Z&=\lim_{r\rightarrow\infty}\left[\frac{\Tr(\mathbb{T}^{r}\mathbb{T}_Z^{r})/\Tr(\mathbb{T}^{2r})}{\sqrt{\Tr(\mathbb{T}_Z^{2r})/\Tr(\mathbb{T}^{2r})}}\right]^{1/2}\notag\\
    &=\left[\lim_{r\rightarrow\infty}\frac{\Tr(\mathcal{T}^{r}\mathcal{T}^r_Z)/\Tr(\mathcal{T}^{2r})}{\sqrt{\Tr(\mathcal{T}_Z^{2r})/\Tr(\mathcal{T}^{2r})}}\right]^{\frac{1}{2}}\label{derive_O_Z}\\
    &=\lim_{r\rightarrow\infty}\frac{t^{\frac{r}{2}}t_Z^{\frac{r}{2}}/t^{r}}{t_Z^{\frac{r}{2}}/t^{\frac{r}{2}}}|\braket{V}{V_Z}|=|\braket{V}{V_Z}|,\label{derive_O_Z_finial}
\end{align}
where we compress the transfer matrices $\mathbb{T}$ and $\mathbb{T}_Z$ to the transfer matrices $\mathcal{T}$ and $\mathcal{T}_Z$ with a dimension $D^2\chi^2$ using the edge tensors from the CTMRG, see Fig.~\ref{Figure_1_prime}b, and $t$ ($t_Z$), $V$ ($V_Z$) are dominant eigenvalue and eigenvector of $\mathcal{T}$ ($\mathcal{T}_Z$). Here, we assume that the dominant eigenvectors are non-degenerate and discuss the degenerate case later.  In the limit of $r\to\infty$, the action of the transfer matrices is set by the dominating eigenvalue and eigenvector, $\mathcal{T}_Z^{\infty}=t_Z^{\infty}\ket{V_Z}\bra{V_Z}$ and $\mathcal{T}^{\infty}=t^{\infty}\ket{V}\bra{V}$, as shown in Fig.~\ref{Figure_1_prime}c, so Eq.~\eqref{derive_O_Z} can be simplified to Eq.~\eqref{derive_O_Z_finial}. Moreover, the perimeter law coefficient can be obtained from the dominant eigenvalues of the transfer matrices: $\alpha_Z=-\log(t_Z/t)$. 

Next, we consider degenerate fixed points of $\mathcal{T}$ and $\mathcal{T}_Z$. This could happen when a ground state in the toric code phase is chosen as the minimally entangled state.
In the limit $r\rightarrow\infty$, $\mathcal{T}^{r}$ ($\mathcal{T}^{r}_Z$) can be expressed as:
\begin{equation}\label{dominant_egv}
  \mathcal{T}^{\infty}=\sum_{\alpha=1}^{d}t^{\infty}\ket{V_{\alpha}}\bra{V_{\alpha}},\quad   \mathcal{T}^{\infty}_Z=\sum_{\alpha_Z=1}^{d_Z}t_{Z}^{\infty}\ket{V_{Z,\alpha_Z}}\bra{V_{Z,\alpha_Z}},
\end{equation}
where $d$ ($d_Z$) denotes the number of the dominant eigenvectors of $\mathcal{T}$ and $\mathcal{T}_Z$, and $\alpha$ ($\alpha_Z$) specifies the degenerate dominant vectors. Substituting Eq.~\eqref{dominant_egv} in Eq.~\eqref{derive_O_Z}, we can calculate the FM string order parameter even when the transfer matrix fixed points are degenerate:
\begin{align}\label{TN_FM_op}
  O_Z=\left[\frac{1}{\sqrt{dd_Z}}\Tr\left(\sum^{d,d_Z}_{\alpha=1,\beta=1}\ket{V_{1,\alpha}}\bra{V_{1,\alpha}}\ket{V_{Z,\beta}}\bra{V_{Z,\beta}}\right)\right]^{\frac{1}{2}}.   
\end{align}


\ \\{\large\textbf{Acknowledgements}}\\ We thank Fengcheng Wu and Youjin Deng for providing their original QMC data used in Ref.~\cite{Youjin_TC_phase_diagram_2011}, and Tibor Rakovszky, Yu-Jie Liu and Rui-Zhen Huang for many helpful comments. We acknowledge support from the Deutsche Forschungsgemeinschaft (DFG, German Research Foundation) under Germany’s Excellence Strategy--EXC--2111--390814868, TRR 360 – 492547816 and DFG grants No. KN1254/1-2, KN1254/2-1, the European Research Council (ERC) under the European Union’s Horizon 2020 research and innovation programme (grant agreement No. 851161 and No. 771537), as well as the Munich Quantum Valley, which is supported by the Bavarian state government with funds from the Hightech Agenda Bayern Plus.

\ \\{\large\textbf{Data and code availability}} \\
Data, data analysis, and simulation codes are available upon reasonable request on Zenodo~\cite{zenodo}.

\ \\{\large\textbf{Note Added}}\\While finalizing the manuscript we became aware of related work on the stability of the FM order parameter~\cite{Verresen2024}.

\ \\{\large\textbf{Author Contributions}}\\
W.-T. X. developed the codes and performed the numerical simulations. All authors contributed to the discussion of the results and the writing of the manuscript.

\ \\{\large\textbf{Competing Interests}}\\The authors declare no competing interests.

 \begingroup
\renewcommand{\addcontentsline}[3]{}
\nolinenumbers
%

\endgroup

\clearpage
\widetext

{\centering
{\large \textbf{ Supplemental Materials for ``Critical behavior of Fredenhagen-Marcu string order parameters\\ at topological phase transitions with emergent higher-form symmetries''}}

\

{Wen-Tao Xu, Frank Pollmann and Michael Knap}

{\em 
{$^1$Department of Physics, Technical University of Munich, 85748 Garching, Germany}

{$^2$Munich Center for Quantum Science and Technology (MCQST), Schellingstr. 4, 80799 M{\"u}nchen, Germany}
}
}

\setcounter{equation}{0}
\setcounter{figure}{0}
\renewcommand{\theequation}{S\arabic{equation}}
\renewcommand{\thefigure}{S\arabic{figure}}

\tableofcontents

\section{Energy density, correlation length, expectation values of local Hamiltonian terms and local order parameters}

\begin{figure}[htp]
    \centering
    \includegraphics[scale=0.5]{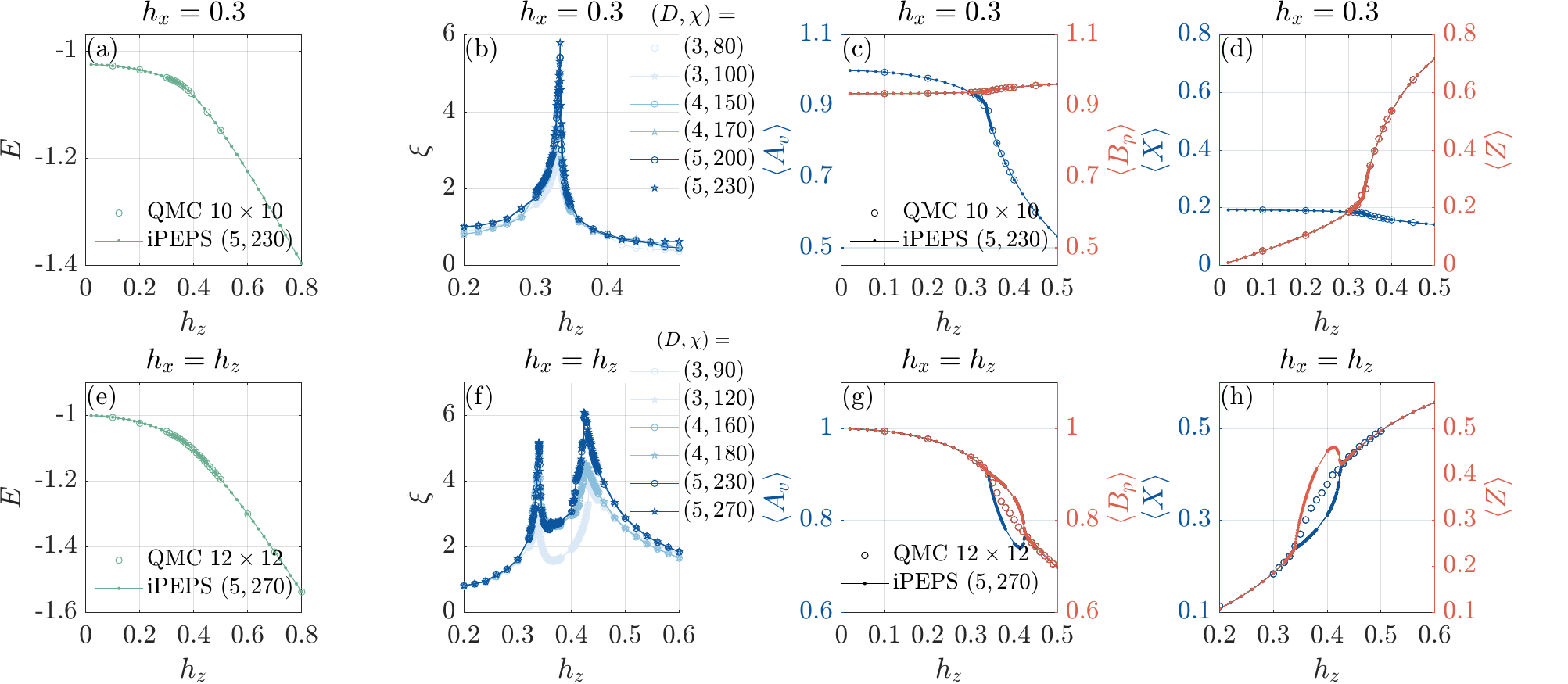}
    \caption{ \textbf{Comparison of ground state energy density and expectation values of local Hamiltonian terms from variational iPEPS and QMC.} (a) Ground state energy density along $h_x=0.3$, the legend shows the bond dimensions $(D,\chi)$ of iPEPS and the system size of the QMC. (b) Correlation length from iPEPS with various bond dimensions along $h_x=0.3$.  (c) Expectation values of $A_v$ and $B_p$ along $h_x=0.3$. (d) Expectation values of $X$ and $Z$ along $h_x=0.3$. (e) Ground state energy density along the self-dual line $h_x=h_z$. (f) Correlation length from iPEPS with various bond dimensions along the self-dual line. (g) Expectation values of $A_v$ and $B_p$ along the self-dual line line. (h) Expectation values of $X$ and $Z$ along the self-dual line.}
    \label{Figure_app_energy_and_expec}
\end{figure}


In this section, we benchmark the ground state energy density and the  expectation values of local terms of the toric code model by comparing the ground state energy density and expectation values of local Hamiltonian terms from our optimized iPEPS and quantum Monte Carlo (QMC) simulations~\cite{Youjin_TC_phase_diagram_2011_app}.
First, we consider the line $h_x=0.3$. The results obtained from iPEPS match perfectly with those from the quantum Monte Carlo (QMC)~\cite{Youjin_TC_phase_diagram_2011_app}, as shown in Figs.~\ref{Figure_app_energy_and_expec}a, c and d. Extrapolating the peak positions of correlation length $h^{(J)}_{zc}(D)$ (see Fig.~\ref{Figure_app_energy_and_expec}b) obtained from the iPEPS with different bond dimensions using a function $h^{(J)}_{zc}(D)=a/D^b-h^{(J)}_{zc}$, where we ignore the $\chi$ dependence and $a, b$ are parameters, we can roughly determine that the phase transition point $J$ shown in Fig.~1a is $h^{(J)}_{zc}=0.335(1)$, which is close to result $h^{(J)}_{zc}=0.333(1)$ of Ref.~\cite{Youjin_TC_phase_diagram_2011_app}.

\begin{figure}
    \centering
    \includegraphics[scale=0.5]{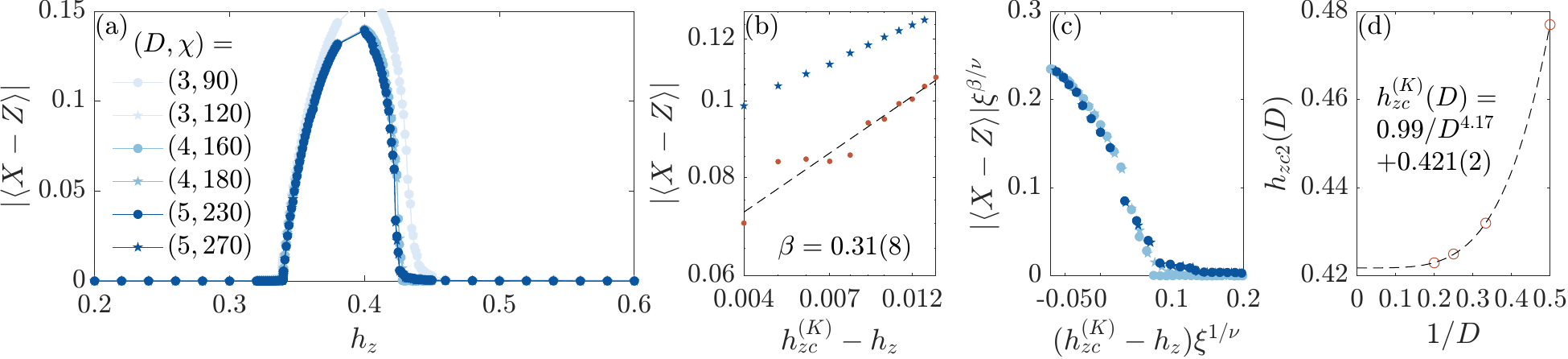}
    \caption{ \textbf{Local order parameter along the self-dual line $(h_x=h_z)$ and its scaling near the critical endpoint $K$ of variational iPEPS.}  (a) Result from iPEPS with various bond dimensions. (b) Double-log plot extracting the critical exponent $\beta_{\text{local}}$ of $|\langle X-Z\rangle|$ by a linearly extrapolation in $1/D$  (red dots), where $h^{(F)}_{zc}=0.421(2)$. The blue stars are $|\langle X-Z\rangle|$ from iPEPS with the bond dimensions $(D,\chi)=(5,270)$. (c) Data collapse of the local order parameter at the critical endpoint $K$, where $\nu=0.629970(4)$ and $\beta=0.326418(2)$. (d) Extrapolating the position of the critical end point $K$ from the peak positions of the correlation length in Fig.~\ref{Figure_app_energy_and_expec}f.}
    \label{Figure_app_local_order_para}
\end{figure}

In Figs.~\ref{Figure_app_energy_and_expec}e, g and h, we compare the energy density and expectation values along the self-dual line $(h_x=h_z)$ with the QMC result. The correlation length shown in the inset of Fig.~\ref{Figure_app_energy_and_expec}f has two peaks corresponding to the multi-critical point $M$ and the critical endpoint $K$.  Different from finite size QMC simulation where the symmetries can not be broken spontaneously, we can obtain $\langle A_v\rangle\neq\langle B_p\rangle$ and  $\langle X\rangle\neq\langle Z\rangle$ for intermediate fields from the iPEPS results, implying a spontaneous duality symmetry breaking. Using the same method for extrapolating the position of $J$, we can roughly determine that the multi-critical point $M$ shown in Fig.~1a is $h^{(M)}_{zc}=0.3397(2)$, which is close to $h^{(M)}_{zc}=0.340(2)$ of Ref.\cite{Youjin_TC_phase_diagram_2011_app} and $h^{(M)}_{zc}=0.3406(4)$ of Ref.~\cite{TC_phase_diagram_expansion_2009_app}. Moreover, the position of the critical endpoint $K$ strongly depends on the bond dimension $D$. As shown in the inset of Fig.~\ref{Figure_app_local_order_para}d, the extrapolated position of $K$ is $h_{zc}^{(K)}=0.421(2)$, which is close to $h_{zc}^{(K)}=0.418(2)$ of Ref.~\cite{Youjin_TC_phase_diagram_2011_app} and indicates that $h_{zc}^{(K)}=0.48(2)$ obtained in Ref.~\cite{TC_phase_diagram_expansion_2009_app} is questionable.


It is natural to expect that the phase transition at the critical endpoint $K$ is also described by the 3D Ising universality class according to the universality hypothesis~\cite{TC_phase_diagram_expansion_2009_app,Self_dual_critically_Ising_app}, because it is a conventional spontaneous $\mathbb{Z}_2$ symmetry breaking phase transition in $(2+1)$D. It is interesting to check the universality hypothesis using the iPEPS simulation results in Fig.~\ref{Figure_app_local_order_para}a. Using the same method for extracting other critical exponents, we obtain $\beta_{\text{local}}=0.31(8)$ defined by $|\langle X-Z\rangle|\sim(h_{zc}^{(K)}-h_z)^{\beta_{\text{local}}}$, as shown in the inset of Fig.~\ref{Figure_app_local_order_para}b; since extrapolated results have large fluctuation, we also show $|\langle X-Z\rangle|$ from iPEPS with bound dimensions $(5,270)$. So $\beta_{\text{local}}=0.31(8)$ is close to $\beta=0.326418(2)$ from the 3D Ising universality class~\cite{Conformal_bootstrap_2016_app}. Moreover, it can be found that the data from $D>3$ can collapse; see Fig.~\ref{Figure_app_local_order_para}c. These results imply that the critical endpoint $K$ is consistent with the 3D Ising universality class.

\section{Discontinuity of the FM string order parameter in the absence of an emergent 1-form symmetry}

We have shown that the FM string order parameter is discontinuous in the confined phase of the deformed toric code wave fucntion, which to some extend is fine tuned. However, it is unclear whether a similar behavior is expected for the variational iPEPS. Directly evaluating the FM string order parameter from the variational iPEPS in the confined phase is not possible due to numerical instability. Instead, we compute a perturbed wavefunction near the infinite large field limit, which does not suffer from numerical instabilities. To this end, we reparameterize the toric code Hamiltonian with $h_x=r\cos(\theta),h_z=r\sin(\theta)$ and apply a unitary transformation
\begin{equation}
    U=\left(\begin{array}{cc}
       \cos (\theta/2)  & \sin (\theta/2)  \\
        -\sin (\theta/2) & \cos (\theta/2) 
    \end{array}\right)
\end{equation}
 to it
\begin{equation}
    U^{\dagger}H_{\tTC}U=-\sum_v A_v'-\sum_p B_p'-r\sum_e X_e,
\end{equation}
where $A'_v=\prod_{e\in v} X_e'$, $B'_p=\prod_{e\in p} Z_e'$, $X'=X\cos\theta -Z\sin \theta$ and $Z'=Z\cos\theta +X\sin \theta$. We first calculate in the $U$-transformed basis and then transform back to the original basis.  
When $r\rightarrow{\infty}$, the ground state of $H_{\tTC}$ can be written as
\begin{equation}
    \ket{\Psi^{[0]}}=\prod_e\ket{\theta}_e,\quad\ket{\theta}=U\ket{+}=\cos\frac{\theta}{2}\ket{+}+\sin\frac{\theta}{2}\ket{-}.
\end{equation}
The FM string order parameter evaluated in this limit is $1$ for $\theta\neq 0$ and $0$ for $\theta=0$.

\begin{figure}
    \centering
    \includegraphics[scale=0.5]{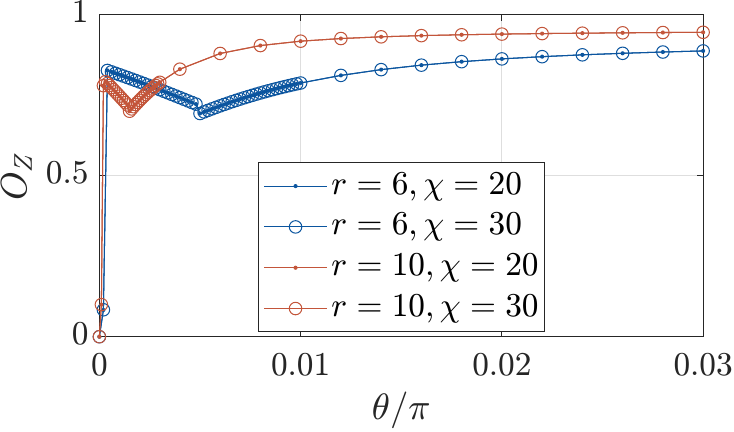}
    \caption{ \textbf{FM string order parameter in the confined region of a perturbatively constructed iPEPS.}  The FM string order parameter evaluated using the iPEPS $\ket{\tilde\Psi^{[1]}}$ in Eq.~\eqref{eq:suppPEPSPert} constructed from first-order perturbation theory around the infinite field limit $r\rightarrow\infty$. $\chi$ is the bond dimension of the CTMRG environment.}
    \label{Figure_app_FM_discontinue}
\end{figure}

Near the infinite field limit, the first order ground state can be written as
\begin{equation}
    \ket{\Psi^{[1]}}=\prod_e U_e\left(1+\sum_v \prod_{e\in v}\sum_{\{\alpha_e=0,1\}}f_v(\{\alpha_e\},r,\theta)Z^{\alpha_i}_e+\sum_p \prod_{e\in p}\sum_{\{\alpha_e=0,1\}}f_p(\{\alpha_e\},r,\theta)Z^{\alpha_i}_e\right)\prod_e\ket{+}_e,
\end{equation}
where
\begin{align}
    f_v(\{\alpha_e\},r,\theta)&=\begin{cases}
     \left[\prod_{e=1}^4(\cos\theta)^{1-\alpha_e}(-\sin\theta)^{\alpha_e}\right]/\left[2r\sum_{e=1}^4\alpha_e\right],&\quad\mbox{if } \sum_{e=1}^4\alpha_e\neq 0;\\
     0,&\quad \mbox{else if }\sum_{e=1}^4\alpha_e=0;
    \end{cases}\\
     f_p(\{\alpha_e\},r,\theta)&=\begin{cases}
     \left[\prod_{e=1}^4(\sin\theta)^{1-\alpha_e}(\cos\theta)^{\alpha_e}\right]/\left[2r\sum_{e=1}^4\alpha_e\right],&\quad\mbox{if } \sum_{e=1}^4\alpha_e\neq 0;\\
     0,& \quad \mbox{else if }\sum_{e=1}^4\alpha_e= 0.
      \end{cases}
\end{align}
We can slightly change $\ket{\Psi^{[1]}}$ to  $\ket{\tilde{\Psi}^{[1]}}$, such that it can be written as an iPEPS, and the difference between $\ket{\tilde{\Psi}^{[1]}}$ and  $\ket{\Psi^{[1]}}$ is $O(1/r^2)$~\cite{Laurens_bridge_2017_app}:  
\begin{align}
    \label{eq:suppPEPSPert}\ket{\tilde{\Psi}^{[1]}}=\prod_eU_e\left\{\prod_v\left[ 1+ \prod_{e\in v}\sum_{\{\alpha_e=0,1\}}f_v(\{\alpha_e\},r,\theta)Z^{\alpha_i}_e\right]\prod_p\left[1+ \prod_{e\in p}\sum_{\{\alpha_e=0,1\}}f_p(\{\alpha_e\},r,\theta)Z^{\alpha_i}_e\right]\right\}\prod_e\ket{+}_e.
\end{align}
Now, $\ket{\tilde{\Psi}^{[1]}}$ can be interpreted as some gates applies on a product state, and it can be easily written as a $2\times2$ unit cell iPEPS with a bond dimension $2$. As a next step, we evaluate the FM string order parameter from the iPEPS of $\ket{\tilde{\Psi}^{[1]}}$; see Fig.~\ref{Figure_app_FM_discontinue}. It is singular near (but not exactly at) the $h_x$ axis and displays non-analytical behavior for some finite $\theta$. This implies that also for the variational iPEPS the FM string order parameter cannot be applied in the absence of an underlying associated 1-form symmetry. 

\section{Dual FM string order parameter and the stability}
Because of the electric-magnetic duality of the toric code model, there is a dual FM string order parameter $O_X$, defined by replacing $Z$ with $X$ as well as the loop (string) on the primal lattice with those on the dual lattice in the FM string order parameter:
\begin{equation}\label{dual_FM_op}
   O_X=\lim_{|\hat{L}_{x,1/2}|\rightarrow\infty}\sqrt{|C_X(|\hat{L}_{x,1/2}|)|}, \quad\quad C_X(|\hat{L}_{x,1/2}|)=\frac{\langle{\Psi}|\prod_{e\in \hat{L}_{x,1/2}}X_e|{\Psi}\rangle/\braket{\Psi}{\Psi}}{\sqrt{\bra{\Psi}\prod_{e\in \hat{L}_x} X_e\ket{\Psi}/\braket{\Psi}{\Psi}}},   
 \end{equation}
  where $\hat{L}_x$ is a non-contractible loop on the dual lattice whose length is twice the length of the string $\hat{L}_{x,1/2}$ on the dual lattice. Along the self-dual line, we expect that both $O_X$ and $O_Z$ are zero. However, the proper ground state needs to be chosen to obtain the desired results. 

\begin{figure}
    \centering
    \includegraphics[width=0.99\linewidth]{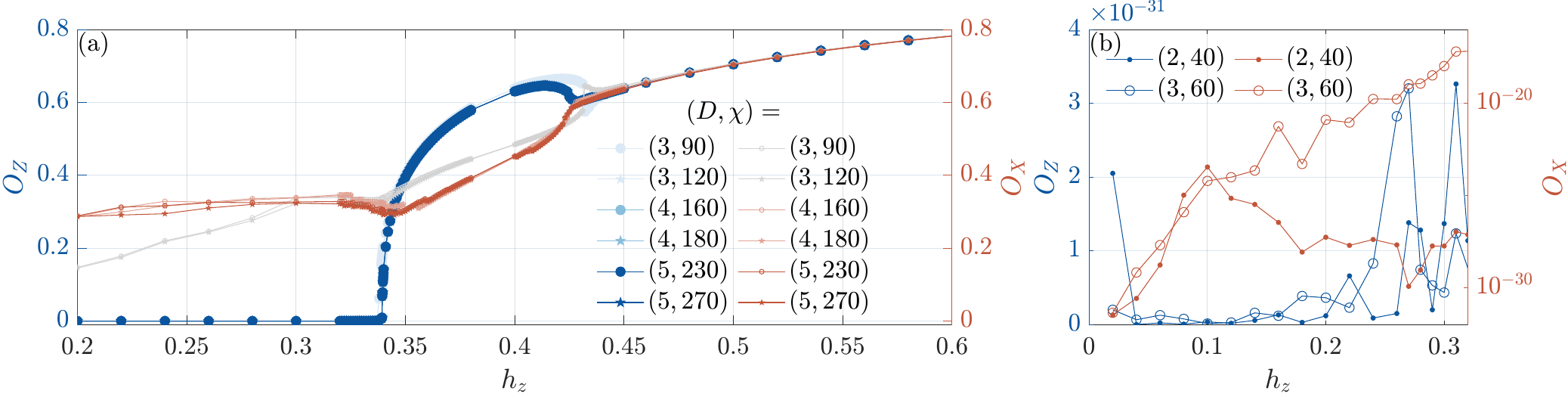}
    \caption{ \textbf{FM string order parameter and its dual of the variational iPEPS.}  (a) Along the self-dual line, the FM string order parameter and its dual from iPEPS that respects the Wilson loop symmetry but spontaneously breaks the 't Hooft loop symmetry with various bond dimensions. The toric code phase, the duality symmetry breaking phase, and the trivial phase arise with increasing field at non-analytic points of the FM order parameter. (b) Along the self-dual line, the FM string order parameter and its dual evaluated using the trivial minimally entangled state in the toric code phase. Both of them are zero (to machine precision). (c) The FM string order parameter $O_Z$ calculated along the line $h_z=0.3$. Its dual $O_X$ along $h_z=0.3$ is equivalent to $O_Z$ along $h_x=0.3$, which is shown in Fig. 2a of the main text. Due to the absence of the $Z$ 1-form symmetry in the confined regime, the FM string order parameter $O_Z$ cannot be used to infer the  properties of the critical point. }
    \label{Figure_app_FM_order_para}
\end{figure}

In Fig.~\ref{Figure_app_FM_order_para}a, we calculate both $O_X$ and $O_Z$ using a ground state $\ket{0_x0_y}$ (the simultaneous ground state of the emergent Wilson operators on the non-contractible loops in $x$ and $y$ directions) that spontaneously breaks the emergent 't Hooft loop symmetry.  It can be found that the $O_Z$ is zero in the toric code phase as expected, but $O_X$ is non-zero and unstable in the toric code phase, indicating that the FM string order parameters are unstable when the corresponding emergent 1-form symmetries are absent.  

If we want to obtain the correct behavior of both $O_X$ and $O_Z$ from the same ground state, we can use the minimally entangled state. At $h_x=h_z=0$, the iPEPS tensor $A$ has a virtual $\mathbb{Z}_2$ symmetry, see Fig. 6a in the main text. The virtual $\mathbb{Z}_2$ symmetry allows us to construct the Wilson loop operators at the virtual level~\cite{Schuch_2011_app}, which are equivalent to the Wilson loop operators on the physical level used to obtain all degenerate ground states. To obtain the minimally entangled states away from the limit $h_x=h_z=0$, we have to impose the virtual $\mathbb{Z}_2$ symmetry to the iPEPS tensor, see Fig.~5b in the main text. We find that the ground state energies obtained from the iPEPS with and without imposing virtual $\mathbb{Z}_2$ symmetry are very close to each other in the toric code phase for various bond dimensions. So, we can safely impose the virtual $\mathbb{Z}_2$ symmetry to the iPEPS tensor in the toric code phase, which corresponds to the emergent Wilson loop symmetry or the emergent 't Hooft loop symmetry on the physical level. We can apply a projector $\lim_{N\rightarrow\infty}(\mathbbm{1}^{\otimes N}_D+Z^{\otimes N}_D)/2$ (see $Z_D$ in Fig.~5b of the main text) to the virtual level of iPEPS to project to the minimally entangled state in the trivial topological sector, see detail in Appendix G of Ref.~\cite{lukas_2023_app}. We find that both $O_X$ and $O_Z$ become zero when the ground state is chosen as the trivial minimally entangled state; see Fig.~\ref{Figure_app_FM_order_para}b. This agrees with the FM string order parameters whose denominators are defined using contractible loops.


In the trivial phase, there are two degenerate ground states in the duality symmetry-breaking phase corresponding to the predominant condensation of charges and fluxes, respectively. In Fig.~\ref{Figure_app_FM_order_para}a, we use the one with predominant condensation of charges, so the FM string order parameter $O_Z$ detecting the charge condensation is well-behaved. However, the dual FM string order parameter $O_X$ detects the flux condensation is discontinuous at the multi-critical point $M$ because $O_X=0$ is in the toric code phase if we use the trivial minimally entangled state. since the charge condensation predominant ground state does not have the emergent 1-form 't Hooft loop symmetry, it implies that the corresponding emergent 1-form symmetries are necessary for the FM string order parameters to be continuous at second-order phase transitions and exhibit proper criticality. 

We further investigate the behavior of $O_Z$ along a path $h_z=0.3$ which crosses the transition between the deconfined phase and the confined regime, where we initialize the iPEPS using the state $\ket{0_x0_y}$ and do not impose the virtual $\mathbb{Z}_2$ symmetry. As show in Fig.~\ref{Figure_app_FM_order_para}c, $O_Z=0$ in the deconfined phase as we expected. However, it becomes unstable in the confined regime because the denominator of the FM string order parameter decays too fast to zero with the length of the loop, and this instability becomes more severe with $h_x$ increasing. Because in the confined regime, in which an underlying $Z$ 1-form symmetry does not exist, the FM string order parameter $O_Z$ cannot capture the critical point between the deconfined phase and the confined regime.  

\section{Equivalence between the FM string order parameters defined using contractible and non-contractible loops}

 In this section, we show that the FM string order parameters defined using a non-contractible loop $L_x$ and evaluated using the trivial minimally entangled state and the one defined using a contractible loop $L$ are equivalent, provided that the strings are infinitely long and the ground state is an iPEPS. First, consider the case with a contractible loop. According to Eq. (3) in the main text, we need to consider the following three infinitely large tensor networks:
\begin{equation}\label{finite_FM_order_parameter}
\vcenter{\hbox{
  \includegraphics{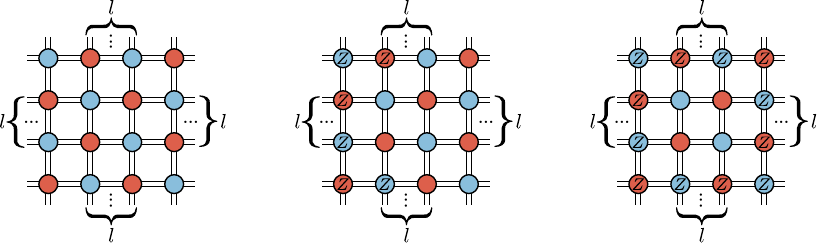}}},
\end{equation}
where the length of the loop and the string are $|L|=4l+4$ and $|L_{1/2}|=2l+2$, respectively. When $l$ is very large, using the edge tensors from the CTMRG, one can replace the middle part with the object shown on the right hand side of the following equation:
\begin{equation}\label{middle_part}
\vcenter{\hbox{
  \includegraphics{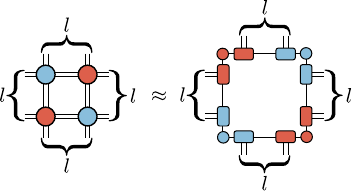}}},
\end{equation}
where the corner tensors (not CTMRG corner tensor) represented by circles can be obtained using a method shown in Ref.~\cite{Laurens_gradient_app}. However, we do not need to calculate the corner tensors represented by circles because they will be canceled later. Since the iPEPS tensors have the virtual $\mathbb{Z}_2$ symmetry shown in Fig.~5b, the right hand side of Eq.~\eqref{middle_part} should also have this symmetry:
\begin{equation}\label{sym_middle_part}
\vcenter{\hbox{
  \includegraphics{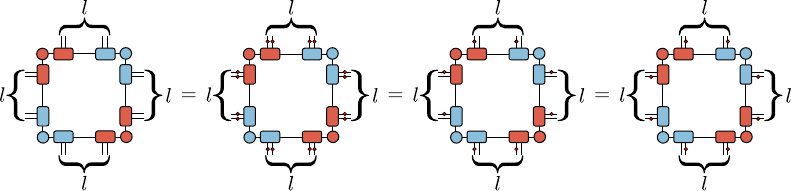}}}.
\end{equation}
Notice it is not always guaranteed that the object on the right side of Eq.~\eqref{middle_part} has the virtual $\mathbb{Z}_2$ symmetry because the  environment of the iPEPS could spontaneously break the virtual $\mathbb{Z}_2$ symmetry~\cite{haegeman2015shadows_app}. If so, we must apply a projector $(\mathbbm{1}^{\otimes N}+Z_D^{\otimes N})/2$ to the object to restore the virtual $\mathbb{Z}_2$ symmetry, similar to what we do for obtaining the trivial minimally entangled state~\cite{lukas_2023_app,xu_entanglement_2023_app}. With the object in the right side of Eq.~\eqref{middle_part} as well as the corner and edge tensors of the CTMRG environment, the three infinity tensor networks in Eq.~\eqref{finite_FM_order_parameter} can be approximated as:
\begin{equation}\label{three_tns_approx}
\vcenter{\hbox{
  \includegraphics{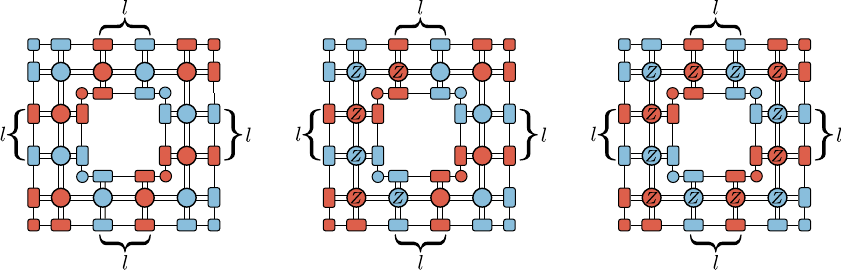}}}.
\end{equation}
Because $l$ is very large, we can replace the power of transfer matrices $\mathcal{T}$ and $\mathcal{T}_Z$ with their fixed points to simplify the three tensor networks in Eq.~\eqref{three_tns_approx}:
\begin{equation}\label{Three_tns_approx_2}
    \vcenter{\hbox{
  \includegraphics{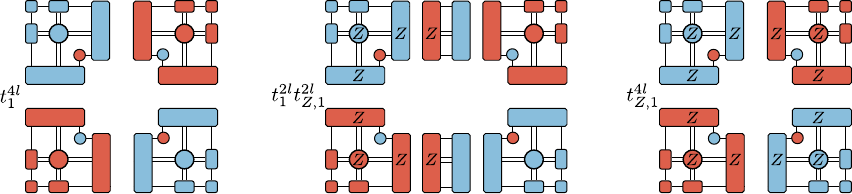}}},
\end{equation}
where we assume that the dominant eigenvectors are not degenerate. We denote the corner objects as 
\begin{equation}\label{corner_object}
    \vcenter{\hbox{
  \includegraphics{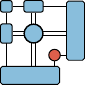}}}= \vcenter{\hbox{
  \includegraphics{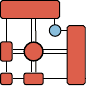}}}=\mathcal{C},  \quad  \vcenter{\hbox{
  \includegraphics{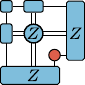}}}= \vcenter{\hbox{
  \includegraphics{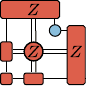}}}=\mathcal{C}_Z,
\end{equation}
where the symmetry of the square lattice is taken into consideration. With Eqs.~\eqref{Three_tns_approx_2} and \eqref{corner_object}, we can express the FM string order parameter as
\begin{equation}
O_Z= \left(\frac{t_1^{2l}t^{2l}_{Z,1}\mathcal{C}^2\mathcal{C}^2_Z|\braket{V_Z}{V}|^2/(\mathcal{C}^4t_1^{4l})}{\sqrt{(\mathcal{C}^4_Zt_{1,Z}^{4l})/(\mathcal{C}^4t_{1}^{4l})}}\right)^{1/2}=\left(\frac{t_1^{2l}t^{2l}_{Z,1}\mathcal{C}^2\mathcal{C}_Z^2|\braket{V_Z}{V}|^2}{\mathcal{C}^2_Z\mathcal{C}^2t_{1,Z}^{2l}t_{1}^{2l}}\right)^{1/2}=|\braket{V_Z}{V}|.
\end{equation}
Comparing with the calculation of the FM string order parameter defined using a non-contractible loop, we can conclude that the FM string order parameters whose denominators are defined using a non-contractible loop and evaluated using the trivial minimally entangled state and the one defined using a contractible loop are equivalent in the limit $|L_{1/2}|\rightarrow \infty$.  When the symmetry depicted in Eq.~\eqref{sym_middle_part} is not satisfied, we should take all four diagrams in Eq.~\eqref{sym_middle_part} separately into account. In the presence of degenerate dominant eigenvectors Eq.~\eqref{Three_tns_approx_2} is not valid. Nonetheless, using the virtual $\mathbb{Z}_2$ symmetry of the iPEPS tensor and the condition shown in Eq.~\eqref{fundmental_MPS}, we can still arrive at the same conclusion. We do not show this more complicated case here.

\section{Analysis of the FM string order parameter for the deformed toric code state}

Here we discuss the relation between the FM string order parameter and the virtual order parameter defined for a topological iPEPS~\cite{Zhu_2019_app} and the origin of the discontinuity of the FM string order parameter without bulk phase transition. 
First, let us review the definition of the virtual order parameter. For a fixed point ground state, we can create a pair of charge excitations at the vertices $v$ and $v^{\prime}$ by inserting two $Z$ operators at the virtual level, because it is equivalent to $Z$ operators along the string $L_{1/2}$ on the physical level:
\begin{equation}\label{Physical2virtual}
    \ket{\pmb{e}_v,\pmb{e}_{v'}}=\vcenter{\hbox{
  \includegraphics{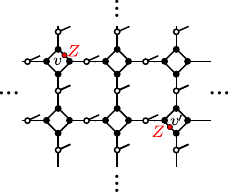}}}= \vcenter{\hbox{
  \includegraphics{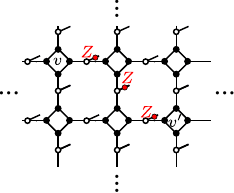}}}.
\end{equation}
Then, the virtual order parameter can be expressed as
\begin{equation}
    O_Z^{(\text{virtual})}=\lim_{|L_{1/2}|\rightarrow\infty}\left(\frac{\bra{\tTC}\prod_{e}Q_e^2(g_x,g_z)\ket{\pmb{e}_v,\pmb{e}_{v'}}}{\bra{\tTC}\prod_{e}Q_e^2(g_x,g_z)\ket{\tTC}}\right)^{1/2}.
\end{equation}
In contrast, the FM string order parameter can be expressed as
\begin{align}\label{FM_order_para_for_deformed_TC}
    O_Z&=\lim_{|L_{L/2}|\rightarrow\infty}\left\{\frac{\bra{\tTC}\left[\prod_{e}Q_e(g_x,g_z)\right]\left(\prod_{e\in L_{1/2}}Z_{e}\right)\left(\prod_{e}Q_{e}(g_x,g_z)\right]\ket{\tTC}}{\sqrt{\bra{\tTC}\left[\prod_{e}Q_e(g_x,g_z)\right]\left(\prod_{e\in L}Z_{e}\right)\left[\prod_{e}Q_{e}(g_x,g_z)\right]\ket{\tTC}}}\right\}^{1/2}\notag\\
   & =\lim_{|L_{1/2}|\rightarrow\infty}\left\{\frac{\bra{\tTC}\left[\prod_{e}Q_e(g_x,g_z)\right]\left[\prod_{e\in L_{1/2}}Q_{e}(g_x,g_z)Q_{e}(-g_x,g_z)\right]\ket{\pmb{e}_v,\pmb{e}_{v'}}}{\sqrt{\bra{\tTC}\left[\prod_{e\notin L}Q^2_e(g_x,g_z)\right]\left[\prod_{e\in L}Q_{e}(g_x,g_z)Q_{e}(-g_x,g_z)\right]\ket{\tTC}}}\right\}^{1/2}.
\end{align}
When $g_x=0$, the FM string order parameter and the virtual order parameter are equal: $O_Z=O^{\text{(virtual)}}_Z$. When $g_x\neq 0$, the FM string order parameter has extra defect lines along the string $L_{1/2}$ in the numerator and the loop $L$ in the denominator compared to the virtual order parameter. If $g_x$ is not too large, the string operator $\prod_{e\in L_{1/2}} Z_e$ will create two charges at its two ends when applied to the deformed toric code state. Thus, the FM string order parameter and the virtual order parameter are quantitatively the same and the FM string order parameter detects the condensation of charges. However, if $g_x$ is too large,  the string operator $\prod_{e\in L_{1/2}} Z_e$ will fail to create the charges when applied to the deformed toric code state. The FM string order parameter becomes essentially different from the virtual order parameter. 

We can understand when the string operator $\prod_{e\in L_{1/2}} Z_e$  fails to create charges from transfer matrices:
\begin{equation}
    \mathcal{T}_{Q^2}=\vcenter{\hbox{
  \includegraphics[width=2.13cm]{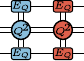}}},\quad \mathcal{T}_{Q\tilde{Q}}=\vcenter{\hbox{
  \includegraphics[width=2.13cm]{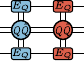}}},
\end{equation}
where
\begin{equation}
 \vcenter{\hbox{
  \includegraphics[width=1.673cm]{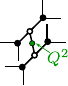}}}=
  \vcenter{\hbox{
  \includegraphics[width=1.066cm]{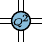}}},\quad\vcenter{\hbox{
  \includegraphics[width=1.673cm]{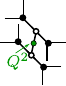}}}=
  \vcenter{\hbox{
  \includegraphics[width=1.066cm]{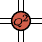}}},\quad
  \vcenter{\hbox{
  \includegraphics[width=1.673cm]{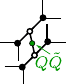}}}=
  \vcenter{\hbox{
  \includegraphics[width=1.066cm]{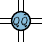}}},\quad\vcenter{\hbox{
  \includegraphics[width=1.673cm]{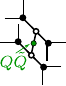}}}=
  \vcenter{\hbox{
  \includegraphics[width=1.066cm]{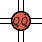}}},
\end{equation}
$\tilde{Q}(g_x,g_z)=-Q(g_x,g_z)$, and $E_Q$ is the boundary MPS tensor of the double tensor labeled by $Q^2$. Notice that the black and white dot tensors are defined in Fig.~6a. Using the transfer matrix fixed points, the virtual order parameter and FM string order parameter can be expressed as
\begin{equation}\label{O_Z_from_fixed_points}
    O_Z^{(\text{virtual})}= \vcenter{\hbox{
  \includegraphics[width=1.494cm]{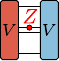}}},\quad O_Z=\vcenter{\hbox{
  \includegraphics[width=1.494cm]{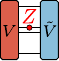}}},
\end{equation}
where $V$ and $\tilde{V}$ are fixed points of $\mathcal{T}_{Q^2}$ and $\mathcal{T}_{Q\tilde{Q}}$, respectively.
Because $\mathcal{T}_{Q^2}$ and $\mathcal{T}_{Q\tilde{Q}}$ has a $\mathbb{Z}_2$ symmetry $U_X\otimes X\otimes X\otimes U_X$ in the toric code phase and the flux condensation phase (this symmetry does not exist in the charge condensation phase because the boundary MPS generated by $E_Q$ spontaneously breaks the virtual $\mathbb{Z}_2$ symmetry~\cite{haegeman2015shadows_app}):
\begin{equation}
   \vcenter{\hbox{
  \includegraphics[width=2.13cm]{T_QQ.pdf}}} = \vcenter{\hbox{
  \includegraphics[width=2.71cm]{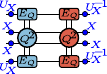}}},\quad  \vcenter{\hbox{
  \includegraphics[width=2.13cm]{T_QQ2.pdf}}}=\vcenter{\hbox{
  \includegraphics[width=2.71cm]{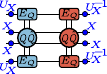}}}.
\end{equation}
where $U_X$ is a $\chi\times\chi$ matrix defined via
\begin{equation}\label{fundmental_MPS}
     \vcenter{\hbox{\includegraphics[width=2.13cm]{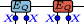}}}=\vcenter{\hbox{
  \includegraphics[width=2.71cm]{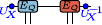}}},
\end{equation}
 we say the parity of $V$ and $\tilde{V}$  is even (odd) if they are eigenstates of $U_X\otimes X\otimes X\otimes U_X$ with eigenvalues 1 (-1). It can be checked that $V$ is always parity even in the toric code phase and the flux condensation phase, so $O^{\text{(virtual)}}_Z=0$ because of Eq.~\eqref{O_Z_from_fixed_points} and  $\{U_X\otimes X\otimes X\otimes U_X,\mathbbm{1}_{\chi}\otimes Z\otimes \mathbbm{1}_2\otimes \mathbbm{1}_{\chi}\}=0$. However, along to $g_x^2+g_z^2=0.65^2$ shown in Figs.~4a and f, $\tilde{V}$ is parity even (odd) when $\theta\gtrsim 0.4\pi$ ($0.25\pi<\theta\lesssim 0.4\pi$), so $O_Z=0$ ($O_Z\neq0$) when $\theta\gtrsim 0.4\pi$ ($0.25\pi<\theta\lesssim 0.4\pi$) according to Eq.~\eqref{O_Z_from_fixed_points}. Therefore, in the flux condensation phase, the abrupt change of parity of $\tilde{V}$ is the origin of the discontinuity of the FM string order parameter in the flux condensation phase.

\begingroup
\renewcommand{\addcontentsline}[3]{}
\nolinenumbers

\endgroup

\begin{thebibliography}{80}
\bibitem [{my_()}]{my_footnote}%
  \BibitemOpen
  \href@noop {} {}\bibinfo {note} {The broken emergent 1-from symmetry defined on a non-contractible loop can be restored using minimally entangled states. Also, notice that the emergent Wilson loop symmetry defined on a contractible loop is unbroken for any ground state.}\BibitemShut {Stop}%
\end{thebibliography}
%

\end{document}